\documentclass{emulateapj}

\usepackage{subfigure}

\def\mybf{}

\def\oldbf{}

\submitted{Accepted for publication in The Astrophysical Journal on 2019 May 15}

\begin{document}
\title{%
A Study of the 20\,Day Superorbital Modulation in the High-Mass X-ray Binary IGR J16493-4348
}%
\author{Joel~B.~Coley\altaffilmark{1,2},
Robin~H.~D.~Corbet\altaffilmark{4,5},
Felix~F\"urst\altaffilmark{6},
Gregory~Huxtable\altaffilmark{4,5},
Hans~A.~Krimm\altaffilmark{3,7},
Aaron~B.~Pearlman\altaffilmark{8,9,10},
Katja~Pottschmidt\altaffilmark{2,4}}
\email{joel.coley@howard.edu}
\altaffiltext{1}{Department of Physics and Astronomy, Howard University, Washington, DC 20059, USA}
\altaffiltext{2}{CRESST/Mail Code 661, Astroparticle Physics Laboratory, NASA Goddard Space Flight Center, Greenbelt, MD 20771, USA}
\altaffiltext{3}{Universities Space Research Association, Columbia, MD 21046, USA}
\altaffiltext{4}{University of Maryland, Baltimore County, MD, USA}
\altaffiltext{5}{CRESST/Mail Code 662, X-ray Astrophysics Laboratory, NASA Goddard Space Flight Center, Greenbelt, MD 20771, USA}
\altaffiltext{6}{European Space Astronomy Center (ESA/ESAC), Operations Department, Vilanueva de la Ca\~nada (Madrid), Spain}
\altaffiltext{7}{National Science Foundation, Alexandria, VA 22314, USA}
\altaffiltext{8}{Division of Physics, Mathematics, and Astronomy, California Institute of Technology, Pasadena, CA 91125, USA}
\altaffiltext{9}{NDSEG Research Fellow}
\altaffiltext{10}{NSF Graduate Research Fellow}



\begin{abstract}

We report on \textsl{Nuclear Spectroscopic Telescope Array (NuSTAR)}, {\mybf \textsl{Neil Gehrels Swift Observatory (Swift)}} X-ray Telescope (XRT) and \textsl{Swift} Burst Alert Telescope (BAT) observations of IGR J16493-4348, a wind-fed Supergiant X-ray Binary (SGXB) showing significant superorbital variability.  From a discrete Fourier transform of the BAT light curve, we refine its superorbital period to be 20.058 $\pm$ 0.007\,days.  The BAT dynamic power spectrum and a fractional root mean square analysis both show strong variations in the amplitude of the superorbital modulation, but no observed changes in the period were found.  The superorbital modulation is significantly weaker between MJD\,55,700 and MJD\,56,300.  The joint \textsl{NuSTAR} and XRT observations, which were performed near the minimum and maximum of one cycle of the 20\,day superorbital modulation, {\oldbf show that the flux increases by more than a factor of two between superorbital minimum and maximum.  We find no significant changes in the 3--50\,keV pulse profiles between superorbital minimum and maximum, which suggests a similar accretion regime.  Modeling the pulse-phase averaged spectra we find a possible Fe K$\alpha$ emission line at 6.4\,keV at superorbital maximum.  The feature is not significant at superorbital minimum.  While we do not observe any significant differences between the pulse-phase averaged spectral continua apart from the overall flux change, we find that the hardness ratio near the broad main peak of the pulse profile increases from superorbital minimum to maximum.  This suggests the spectral shape hardens with increasing luminosity.  We discuss different mechanisms that might drive the observed superorbital modulation.}

\end{abstract}

\section{Introduction}
\label{Introduction}

IGR J16493-4348 is a High-Mass X-ray Binary (HMXB) first discovered during a survey of the Galactic plane using the \textsl{INTErnational Gamma-Ray Astrophysics Laboratory} \citep[\textsl{INTEGRAL};][]{2003A&A...411L...1W} satellite \citep{2004ApJ...607L..33B}.  During a deep scan of the Norma Arm region using \textsl{INTEGRAL}, it was later identified by \citet{2005ATel..457....1G} to be a variable source with a mean photon flux of 5.6 $\pm$ 0.6\,mCrab in the 18--45\,keV energy band.  Two pointed observations using the \textsl{Rossi X-ray Timing Explorer (RXTE)} Proportional Counter Array (PCA) revealed the mean X-ray spectrum to be consistent with a highly absorbed power law.  Its photon index and neutral hydrogen absorbing column were found to be 1.4 and $\sim$10$^{23}$\,cm$^{-2}$, respectively \citep{2005ATel..465....1M}.

A spectral analysis using the {\mybf \textsl{Neil Gehrels Swift Observatory (Swift)}} Burst Alert Telescope (BAT) and the \textsl{INTEGRAL} Soft Gamma-ray Imager (ISGRI), together with pointed \textsl{Swift} {\mybf X-ray Telescope} XRT and \textsl{Suzaku} observations, revealed a hint of an absorption feature at 33 $\pm$ 4\,keV thought to be a Cyclotron Resonant Scattering Feature (CRSF), implying a magnetic field of (3.7 $\pm$ 0.4)$\times$10$^{12}$\,G \citep{DAi2011}.  The width of the absorption feature was found to be 10\,keV.

A single source in the Two Micron All-Sky Survey (2MASS) catalog, 2MASS J16402695-4349090, was identified as the infrared counterpart \citep{2005ATel..654....1K}.  Using the European Southern Observatory (ESO) Infrared Spectrometer and Array Camera (ISAAC) spectrograph on UT1 at Paranal observatory, \citet{2010A&A...516A.106N} proposed the spectral type of the donor star to be B0.5--1 Ia--Ib.  The distance to the source was estimated to be between {\mybf 6 and 26}\,kpc \citep{2010A&A...516A.106N}, but could not be tightly constrained due to the uncertainty of the intrinsic colors.

A $\sim$6.78\,day orbital period was independently found by \citet{2010ATel.2599....1C} and \citet{2010A&A...510A..48C}.  It was later refined by \citet{2013ApJ...778...45C} to be 6.782 $\pm$ 0.001\,days using the BAT Transient Monitor.  The neutron star is regularly eclipsed by the donor star for about 0.8\,days of every orbit, which indicates the orbital inclination is close to edge-on \citep{2019ApJ...873...86P}.  From an eclipse timing analysis using the \textsl{Swift} BAT and \textsl{RXTE} PCA, \citet{2019ApJ...873...86P} further refined the orbital period to be 6.7828 $\pm$ 0.0004\,days.

Recently, \citet{2019ApJ...873...86P} placed constraints on the nature of the donor star using their eclipse timing results.  They proposed the spectral type of the donor star and the distance to the source to be B0.5 Ia and {\oldbf 16.1\,$\pm$1.5\,kpc}, respectively.  We adopt these measurements in this work.

\citet{2010ATel.2766....1C} found evidence of a $\sim$1069\,s signal using the \textsl{RXTE} PCA, which they interpreted as the neutron star rotation period.  From a pulsar timing analysis using an extended PCA dataset, this was later refined to 1093.1036 $\pm$ 0.0004\,s \citep{2019ApJ...873...86P}.  The epoch of maximum delay time, $T_{\pi/2}$, and pulse period derivative were found to be MJD\,55,850.91 $\pm$ 0.05 and 5.4$^{+7.9}_{-9.7}$$\times$10$^{-8}$\,s s$^{-1}$, respectively \citep{2019ApJ...873...86P}.

In addition to the neutron star rotation and orbital periods, a longer superorbital period was observed from IGR J16493-4348.  Using data from the \textsl{Swift} BAT 58 month survey and the \textsl{RXTE} Galactic plane scans, a $\sim$20\,day modulation was found \citep{2010ATel.2599....1C}.  The superorbital period was later refined to be 20.07 $\pm$ 0.01\,days using the BAT Transient Monitor \citep{2013ApJ...778...45C}.  More recently, \citet{2019ApJ...873...86P} refined the superorbital period to be 20.067 $\pm$ 0.009\,days also using the BAT.

Superorbital modulation was additionally seen in the wind-fed {\mybf Supergiant X-ray Binaries (SGXBs)} 2S 0114+650 \citep{2008MNRAS.389..608F}, IGR J16418-4532, IGR J16479-4514, and 4U 1909+07 \citep{2013ApJ...778...45C}. {\oldbf More recently, \citet{2018ATel11918....1C} reported evidence of superorbital modulation in 4U 1538-522.}  In their review of wind-fed SGXBs showing strong superorbital modulation, \citet{2013ApJ...778...45C} found a possible correlation between the orbital and superorbital periods of these binaries, but the mechanism to account for this correlation remains unclear.  Superorbital variability in wind-fed SGXBs is not a ubiquitous feature since many wind-fed SGXBs show strong orbital modulation but no signs of superorbital modulation \citep{2013ApJ...778...45C}.

In Roche-lobe overflow systems, superorbital variations can typically be explained by X-ray irradiation from a central source illuminating a tilted and/or warped accretion disc, causing it to precess and periodically obscure the compact object from the line of sight \citep{1996MNRAS.281..357P,2001MNRAS.320..485O}.  {\mybf Similar variability has also recently been found in ultraluminous X-ray pulsars \citep[e.g. NGC 5907 ULX1; NGC 7793 P13; M82 X-2,][]{2016ApJ...827L..13W,2018A&A...616A.186F,2019ApJ...873..115B}.}  However, the mechanism responsible for the long timescale modulation in wind-fed SGXBs remains poorly understood.  {\mybf Depending on the angular momentum transferred to the compact object by the stellar wind, accretion in wind-fed SGXBs may be mediated by a quasi-spherical outflow \citep{1944MNRAS.104..273B} or by an accretion disk-like structure albeit of a transient nature \citep{2018arXiv181012933E,2018arXiv180805345T,2019RAA....19...12T}.  Indeed, transient accretion disks have been observed in some wind-fed SGXBs \citep[e.g. OAO 1657-415; 2S 0114+650,][]{2012ApJ...759..124J,2017ApJ...844...16H}.}  While {\mybf it is unlikely that a precessing warped and/or tilted accretion disk is the primary mechanism that drives superorbital modulation in wind-fed SGXBs, therefore, it is possible that the superorbital variations are caused by} variable mass accretion rate.  Possible mechanisms that could drive superorbital variations in wind-fed SGXBs include neutron star precession \citep{2013MNRAS.435.1147P}, donor star variability \citep{2006A&A...458..513K}, or the presence of a third star in a hierarchical system \citep{2001ApJ...563..934C}.  Recently, \citet{2017A&A...606L..10B} proposed that a corotation interaction region with a period of $\sim$10.3\,days could explain the $\sim$20.07 day superorbital period and amplitude in IGR J16493-4348.

In this paper, we analyze two \textsl{NuSTAR} and \textsl{Swift} XRT observations of IGR J16493-4348 near the maximum and the minimum of one cycle of the $\sim$20\,day superorbital modulation, together with \textsl{Swift} BAT Transient Monitor observations, which track the evolution of the superorbital modulation on long timescales.  The remainder of the paper is organized as follows. \textsl{NuSTAR} and \textsl{Swift} observations are presented in Section~\ref{Data Analysis and Modeling}.  Section~\ref{Long-Term Variability} focuses on long-term monitoring of the $\sim$20\,day superorbital modulation with the \textsl{Swift} BAT.  In Section~\ref{Temporal Analysis}, we measure the neutron star rotation period using the \textsl{NuSTAR} X-ray telescope and show pulse profiles and their energy dependence at superorbital minimum and superorbital maximum.  Pulse phase averaged and phase resolved spectral results are given in Sections~\ref{Phase-Averaged Spectral Analysis} and~\ref{Phase-Resolved Spectral Analysis}, respectively.  Section~\ref{Pulse-peak Spectral Analysis} focuses on the spectroscopy at the peak of the pulse profile.  We provide a discussion of the results in Section~\ref{Discussion} and the conclusions are given in Section~\ref{Conclusion}.  If not stated otherwise, the uncertainties and limits presented in the paper are at the 90$\%$ confidence level.

\section{Data and Analysis}
\label{Data Analysis and Modeling}

The observations outlined below consist of nearly simultaneous \textsl{NuSTAR} and \textsl{Swift} XRT observations during superorbital minimum (2015 Aug. 31--Sep. 1) and superorbital maximum (2015 Sep. 12), as well as long-term observations of the system with the \textsl{Swift} BAT.  An observation log is given in Table~\ref{X-ray Observations Summary}.

\begin{deluxetable*}{cccccccc}
\tablecolumns{8}
\tabletypesize{\small}
\tablewidth{0pc}
\tablecaption{Summary of X-ray Observations of IGR J16493-4348}
\tablehead{
\colhead{Obs.} & \colhead{Telescope} & \colhead{ObsID} & \colhead{Start Time} & \colhead{End Time} & \colhead{Orbital Phase$^a$} & \colhead{Superorbital Phase$^b$} & \colhead{Exposure} \\
\colhead{} & \colhead{} & \colhead{} & \colhead{(UT)} & \colhead{(UT)} & \colhead{} & \colhead{} & \colhead{(ks)}}
\startdata
Min & \textsl{NuSTAR} & 30102054004 & 2015-08-31 07:23:41 & 2015-09-01 00:26:38 & 0.527--0.632 & 0.520--0.555 & 31.2 \\
Max & \textsl{NuSTAR} & 30102054006 & 2015-09-12 04:40:20 & 2015-09-12 15:17:55 & 0.280--0.344 & 1.113--1.135 & 21.6 \\
\tableline
Min & \textsl{Swift} & 00081665002 & 2015-08-31 10:25:51 & 2015-08-31 10:34:38 & 0.545--0.546 & 0.526--0.527 & 0.5$^c$ \\
Max & \textsl{Swift} & 00081665003 & 2015-09-12 13:00:01 & 2015-09-12 14:56:54 & 0.331--0.343 & 1.130--1.134 & 1.9$^c$ \\
\enddata
\tablecomments{
$^a$ Orbital phase zero is defined at MJD\,55,851.2, corresponding to the epoch of maximum delay time, $T_{\pi/2}$ \citep{2019ApJ...873...86P}. \\*
$^b$ Superorbital phase zero is defined as the epoch of maximum flux (MJD\,57,254.9 $\pm$ 0.3). \\*
$^c$ Net exposure time is spread over several snapshots.}
\label{X-ray Observations Summary}
\end{deluxetable*}

\subsection{\textsl{NuSTAR} Observations}
\label{NuSTAR description}

\textsl{NuSTAR} \citep{2013ApJ...770..103H} carries two co-aligned grazing incidence Wolter I imaging telescopes that focus X-rays between 3--79\,keV onto two independent solid state Focal Plane Modules (hereafter FPMA and FPMB).  We reduced and screened the data using the \textsl{NuSTAR} Data Analysis Software (NuSTARDAS) v.1.7.0 package provided under HEAsoft v.6.20 and calibration files dated 2016 December 07.  The data were reprocessed with the NuSTARDAS data pipeline package \texttt{nupipeline} using the standard filtering procedure to apply the newest calibration and default screening criteria.

The source spectra were extracted in mode 01 (\texttt{SCIENCE}) from a circular region of radius 60$\arcsec$.0 centered on the source.  Since the Norma arm is a crowded region, we checked for stray light contamination produced by sources outside the field of view using the scripts made available on the \textsl{NuSTAR} GitHub webpage\footnote{https://github.com/NuSTAR.}.  We found that FPMA is affected by stray light from multiple sources.  To investigate variations in the background due to stray light, we tested different background regions on the same detector as the source while avoiding visible stray light. We found the spectral parameters do not significantly depend on the choice of background (see Sections~\ref{Phase-Averaged Spectral Analysis}--~\ref{Pulse-peak Spectral Analysis}).  This is not surprising since IGR J16493-4348 was found to be a factor of 10 times brighter than the background at energies below 30\,keV and a factor of two at energies above 30\,keV.  We therefore chose to extract a background from a circular region of radius 60$\arcsec$.0 offset from the source.  Event times were corrected to the solar system barycenter using \texttt{nuproducts} and the FTOOL \texttt{barycorr} with the DE-200 solar system ephemeris.  For the timing analysis and pulse-phase resolved spectra, we further corrected the event times for the orbital motion of the neutron star using the ephemeris defined in \citet{2019ApJ...873...86P}, which assumed a circular orbital solution and no change in the neutron star rotation period {\oldbf (see Section~\ref{Introduction})}.  For phase resolved spectra, Good Time Intervals (GTIs) were generated using the \texttt{nuproducts} tool and the ``usrgtifile" keyword.  Response matrices were generated using the packages \texttt{numkarf} and \texttt{numkrmf}.

The net count rates from the source over the full energy range were found to be 0.595 $\pm$ 0.004 counts s$^{-1}$ (FPMA) and 0.571 $\pm$ 0.004\,counts s$^{-1}$ (FPMB) at superorbital minimum and 1.679 $\pm$ 0.009\,counts s$^{-1}$ (FPMA) and 1.651 $\pm$ 0.009 counts s$^{-1}$ (FPMB) at superorbital maximum.  The background was found to dominate at energies exceeding $\sim$40\,keV at superorbital minimum and $\sim$50\,keV at superorbital maximum.  As a result, we chose to analyze the spectra between 3--40\,keV.  We rebinned the spectral file produced by \texttt{nuproducts} to have a minimum of 50\,counts per bin using \texttt{grppha}.

\subsection{Swift}

\subsubsection{XRT Observations}
\label{XRT description}

{\oldbf The \textsl{Swift} XRT \citep{2005SSRv..120..165B} is a Wolter I imaging telescope sensitive to X-rays ranging from 0.3 to 10 keV.}  We reduced and screened the data using the HEAsoft v.6.20 package and calibration files dated 2017 May 1, following the procedures defined in the XRT Data Reduction Guide (Capalbi et al. 2005).  The data were reprocessed with the XRTDAS standard data pipeline package \texttt{xrtpipeline} using the standard filtering procedure to apply the newest calibration and default screening criteria.  All data were taken in Photon Counting \citep[PC;][]{2004SPIE.5165..217H} mode with a data readout time of 2.5\,s, adopting the standard grade filtering (0--12 for PC).

We find the non-background subtracted count rates at superorbital minimum and superorbital maximum to be {\oldbf 0.15 $\pm$ 0.01}\,counts s$^{-1}$ and {\oldbf 0.25 $\pm$ 0.01}\,counts s$^{-1}$, respectively.  Since our observations of IGR J16493-4348 were not affected by pile-up, we extracted the source spectra from circular regions of {\oldbf radius} 30$^{\prime \prime}$ centered on the source.  The backgrounds were extracted from an annular region of internal radius 60$^{\prime \prime}$ and external radius 120$^{\prime \prime}$ centered on the source.  The ancillary response files, accounting for vignetting, point-spread function correction, and different extraction regions, were generated and corrected for exposure using the FTOOL packages \texttt{xrtmkarf} and \texttt{xrtexpomap}, respectively. 

We further processed the spectral data produced by \texttt{xselect} using the FTOOL \texttt{grppha}, which defined the binning and quality flags of the spectra.  We used the quality flag to further eliminate bad data.  Bins were grouped to ensure a minimum of 20\,counts per bin.

\subsubsection{BAT Observations}

The BAT, on board the \textsl{Swift} spacecraft, is a hard X-ray telescope operating in the 14--195\,keV energy band \citep{2005SSRv..120..143B}.  It provides an all-sky hard X-ray survey with a sensitivity of $\sim$1\,mCrab \citep{2010ApJS..186..378T}.  We analyzed BAT data obtained during the time period MJD\,53,416--57,923 (2005 February 15--2017 June 19).  Light curves were retrieved using the extraction of the BAT Transient Monitor data available on the NASA GSFC HEASARC website\footnote{http://heasarc.gsfc.nasa.gov/docs/swift/results/transients/} \citep{2013ApJS..209...14K}.  We used the orbital light curves in the 15--50\,keV energy band in our analysis, which have exposures that range from 64\,s to 2640\,s in each time bin (see Section~\ref{Long-Term Variability}).  The mean exposure in the time bins is 706\,s.  The short exposures can arise due to the observing plan of \textsl{Swift} since the BAT is primarily tasked to observe gamma-ray bursts \citep{2013ApJS..209...14K}.

The light curves were further screened to exclude bad quality points.  We only considered data where the data quality flag (``DATA\_FLAG") was set to 0, indicating good quality.  Data flagged as ``good" are sometimes suspect, where a small number of data points with very low fluxes and implausibly small uncertainties were found \citep{2013ApJ...778...45C}.  These points were removed from the light curves.  We corrected the photon arrival times to the solar system barycenter using the scripts made available on the Ohio State Astronomy webpage\footnote{http://astroutils.astronomy.ohio-state.edu/time/}.

\section{Results}
\label{Results}

\subsection{Long-Term Variability}
\label{Long-Term Variability}

\begin{figure}[h]
\centerline{\includegraphics[width=3.3in]{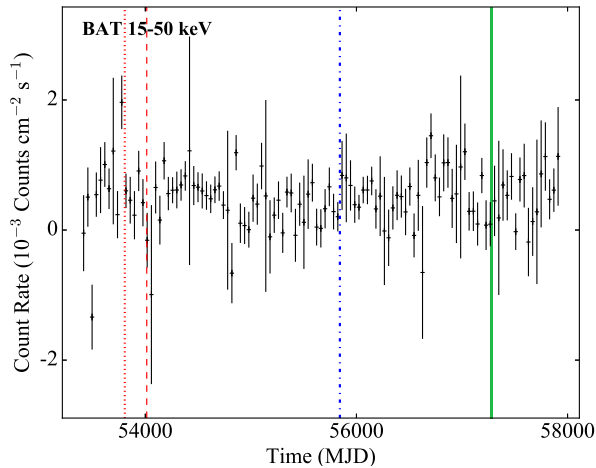}}
\figcaption[June252018_BATrebinned.eps]{
Long-term \textsl{Swift} BAT light curve of IGR J16493-4348 in the 15--50\,keV band (MJD\,53,416--57,923). The light curve is binned to a time resolution chosen to be two superorbital cycles ($\sim$40.13\,days).  The superorbital cycle coinciding with the times of the \textsl{NuSTAR} observations is indicated by the green shaded region.  The \textsl{Swift} XRT,  \textsl{Suzaku}, and \textsl{RXTE} PCA observations reported in \citet{2008MNRAS.385..423H},  \citet{2009ApJ...699..892M} and \citet{2019ApJ...873...86P}, respectively, are indicated by the dotted red, dashed red and dotted-dashed blue lines, respectively.
\label{BAT Transient Monitor}
}
\end{figure}

The \textsl{Swift} BAT Transient Monitor light curve of IGR J16493-4348 is shown in Figure~\ref{BAT Transient Monitor}.  We rebinned the light curve to two superorbital cycles and found no major variability over the duration of the light curve.

\begin{figure*}[h]
\centerline{\includegraphics[width=4.5in]{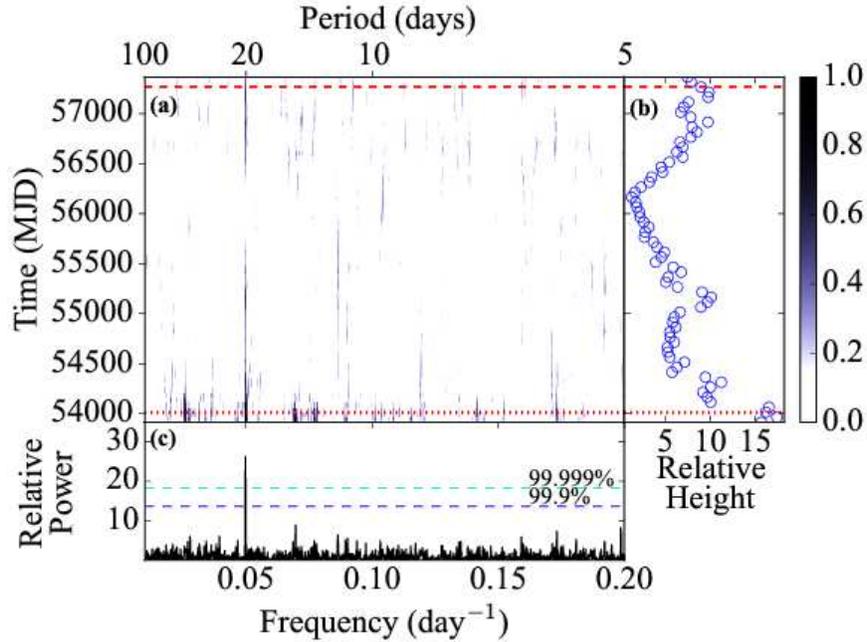}}
\figcaption[March18-2019_1000day-DPSDshrink.eps]{
(a) \textsl{Swift} BAT dynamic power spectrum in the 15--50\,keV band of IGR J16493-4348 as a function of time.  The power spectra were calculated using 1000\,day time intervals, with 50\,day increments in the start and end times.  The superorbital cycle coinciding with the times of the \textsl{NuSTAR} observations near superorbital minimum and maximum is indicated by the red dashed line.  The \textsl{Suzaku} observation reported in \citet{2009ApJ...699..892M} is indicated by the dotted red line.  (b) Relative height of the peak near the $\sim$20.06\,day superorbital period in the power spectrum for each 1000\,day time interval. (c) DFT of the entire data set produced, with 99.9$\%$ and 99.999$\%$ significance levels indicated by the blue and green dashed lines, respectively.
\label{Dynamic Power Spectrum}
}
\end{figure*}

We derived the superorbital period using a Discrete Fourier transform (DFT) of the BAT light curve, after removing points that fell within the eclipse of the neutron star from the start of ingress to the end of egress using the ephemeris defined in \citet{2019ApJ...873...86P}.  The DFT covered a period range between 0.07\,days and the length of the light curve -- i.e. $\sim$4507\,days.  We weighted the contribution of each data point by its uncertainty when calculating the power spectrum using the ``semi-weighting" technique \citep{2007PThPS.169..200C, 2013ApJ...778...45C}, where the error bars on each data point and the excess variability of the light curve are taken into account.  The significance of the peak at the superorbital period was estimated using the false-alarm probability \citep[FAP;][]{1982ApJ...263..835S}, which depends on the number of independent frequencies and therefore the nominal frequency resolution.  While this is not precisely defined for unevenly sampled data \citep{1990ApJ...348..700K}, the inverse of the light-curve length provides a reasonable approximation \citep{2017ApJ...846..161C}.  The uncertainty in our period measurements is obtained using the expression given in \citet{1986ApJ...302..757H}.

The $\sim$20\,day superorbital modulation is strongly detected in the discrete Fourier transform of the BAT Transient Monitor light curve.  Using an additional 674\,days of data compared to \citet{2019ApJ...873...86P}, we refine the superorbital period to 20.058 $\pm$ 0.007\,days (see Figure~\ref{Dynamic Power Spectrum}{\oldbf (c)}). The FAP is 3$\times$10$^{-7}$.  We note that by excising the eclipses, gaps with a spacing of about 1.7\,days are created in the light curve, which could possibly lead to aliasing effects in the power spectrum.  To investigate this, we created a light curve using the times of the BAT light and replaced the data values with a sinusoidal modulation at 20.058\,days.  We find no evidence of aliasing in the power spectrum.

To monitor changes in the $\sim$20.06\,day modulation, we constructed dynamic power spectra using the \textsl{Swift} BAT light curve (see Figure~\ref{Dynamic Power Spectrum}{\oldbf (a)}).  To investigate whether changes in the signal were sudden or gradual, overlapping light curve subsets were used \citep[e.g.][]{2003MNRAS.339..447C}.  We divided the light curve into 70 data windows, where each had a length of 1000\,days{\oldbf , that were successively shifted in time by 50\,days} relative to each other.  We calculated the DFT from each subset of data.  In Figure~\ref{Dynamic Power Spectrum}(a), we show that the amplitude of the $\sim$20.06\,day modulation changes as a function of time.  We find no change in the period of the $\sim$20.06\,day modulation.

In Figure~\ref{Dynamic Power Spectrum}(b), we show changes in the strength of the $\sim$20.06\,day modulation relative to the {\oldbf average value for each one of the individual 70 power spectra.}  We find the peak power to be more than ten times the mean power up to $\sim$MJD\,54,300 and again at MJD\,55,000--55,200 (2009 September 26--2010 January 4).  The relative peak height is found to be near constant at five times the mean power level between MJD\,54,300--55,000.  From MJD\,55,500 to MJD\,56,100 (2012 June 22), we find that it decreases reaching a minimum of 1.3 times the mean power level.  The power then increases linearly up to MJD\,56,400 (2013 April 18) where it is again larger than five times the mean power level.

\begin{figure*}[h]
\centerline{\includegraphics[width=4.5in]{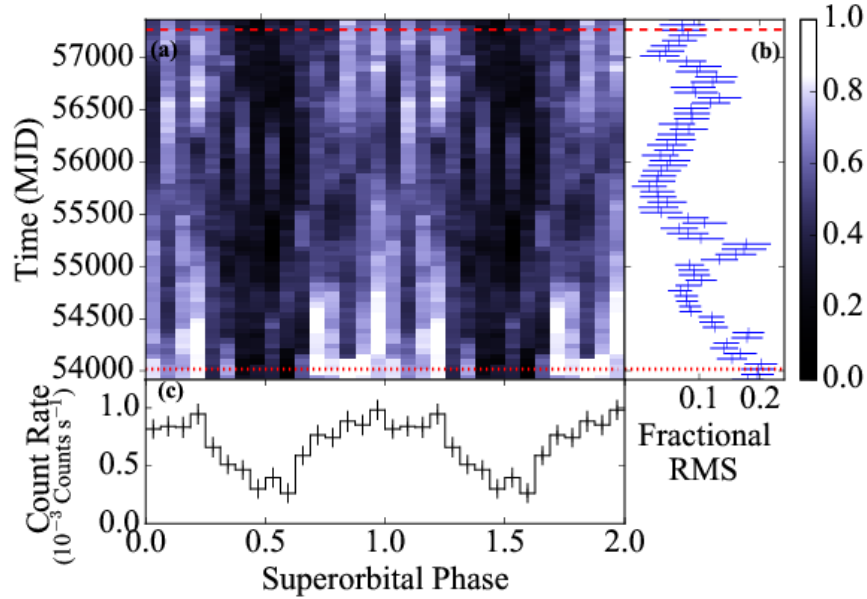}}
\figcaption[March7-2019_1000day-dfoldshrink.eps]{
a) \textsl{Swift} BAT light curve folded on the $\sim$20.06\,day superorbital period using 16 bins as a function of time (see text for details).  b) The variation in the fractional root mean square amplitude of the modulation as a function of time.  The dynamic folded light curves and fractional rms were calculated for 1000\,day time intervals, with 50\,day increments in the start and end times.  The superorbital cycle coinciding with the times of the \textsl{NuSTAR} observations is indicated by the short red dashed line.  The \textsl{Suzaku} observation reported on in \citet{2009ApJ...699..892M} is indicated by the dotted red line. (c) \textsl{Swift} BAT light curve folded on the $\sim$20.06\,day superorbital period using 16 bins.
\label{Dynamic Folded Light Curve}
}
\end{figure*}

Quasi-sinusoidal behavior was found in the BAT Transient Monitor light curve folded on the superorbital period (see Figure~\ref{Dynamic Folded Light Curve}(c)) where we defined phase zero as the epoch of the maximum flux derived from a sine wave fit.  Since the \textsl{NuSTAR} observation near superorbital minimum began at MJD\,57,265.3 (see Table~\ref{X-ray Observations Summary}), we express the epoch of maximum flux (MJD\,57,254.9 $\pm$ 0.3) at an epoch closest to the \textsl{NuSTAR} observation assuming no appreciable change in the superorbital period.

We investigate changes in the amplitude and phase of the superorbital modulation using a dynamic folded light curve (see Figure~\ref{Dynamic Folded Light Curve}(a)).  We divided the light curve into 70 data windows, which each had a length of 1000\,days and were shifted in time by 50\,days relative to each other.  We folded the light curve from each subset of data on the 20.058 $\pm$ 0.007\,day period.  The dynamic folded light curve shows a maximum and minimum near superorbital phases $\sim$0.9--0.2 and $\sim$0.4--0.6, respectively.

To further investigate changes in the amplitude of the modulation, we calculated the fractional root mean square (rms) amplitude and its uncertainty for each 1000\,day segment using Equations 10 and B2 in \citet{2003MNRAS.345.1271V}, respectively (see Figure~\ref{Dynamic Folded Light Curve}{\oldbf (b)}). We find the fractional rms amplitude to track the power of the $\sim$20.06\,day modulation as a function of time.  The fractional rms analysis shows the amplitude of the superorbital modulation significantly decreased to less than 3$\%$ between MJD\,55,700 and MJD\,56,300, which is consistent with the weakening in the dynamic power spectrum.  The weighted Pearson correlation coefficient between the fractional rms and relative height is found to be $r=$0.83, {\mybf with a probability arising by chance of 5$\times$10$^{-8}$.}

\subsection{Short-Term Temporal Analysis}
\label{Temporal Analysis}

\begin{figure*}[t]
    \centering
    \begin{tabular}{cc}
    \subfigure
    {
        \includegraphics[trim=0cm 0cm 0cm 0cm, clip=false, scale=0.4, angle=0]{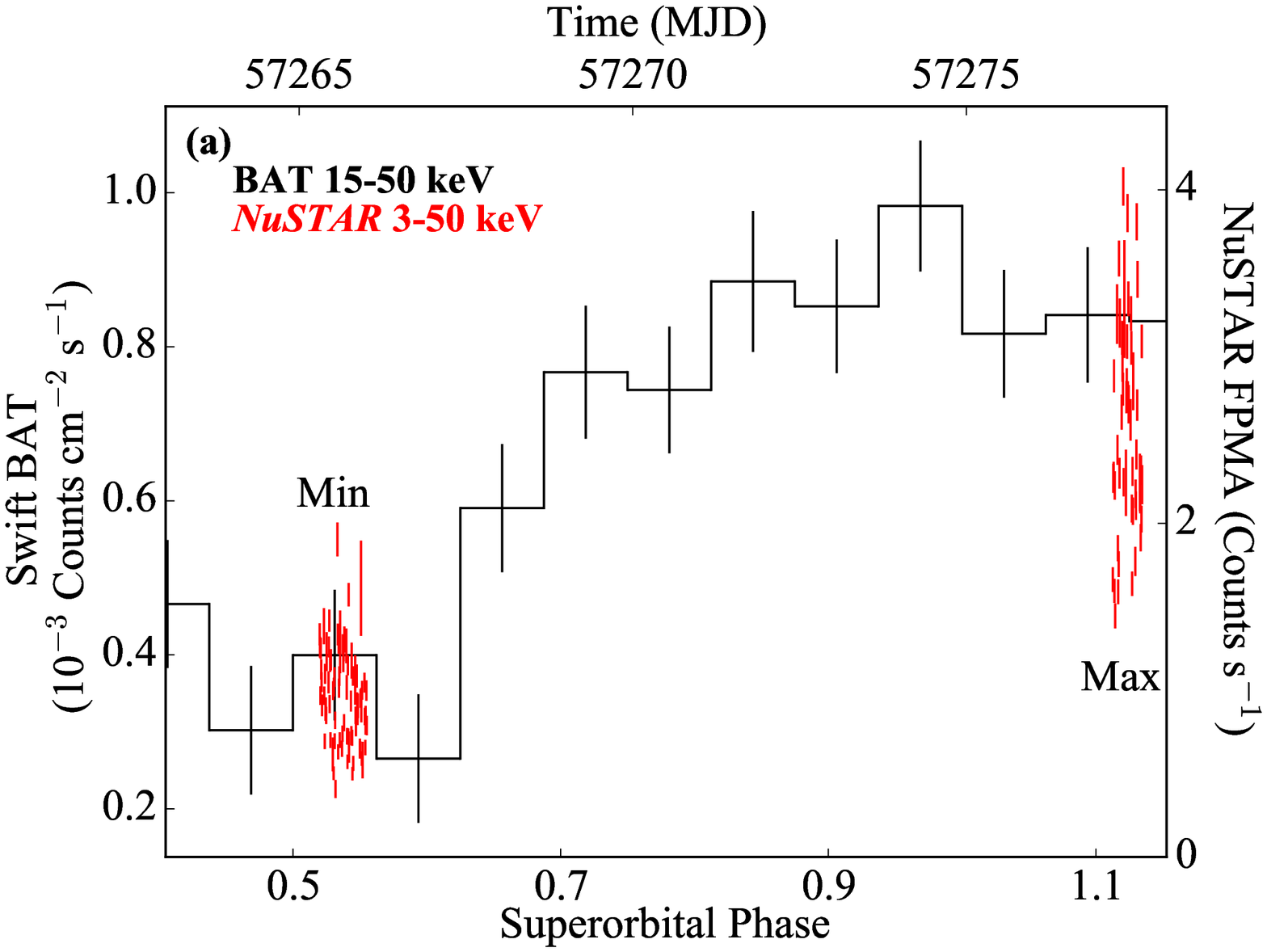}
        \label{Figure:July15-2018_SuperBATNuSTAR.eps}
    }
    &
    \subfigure
    {
        \includegraphics[trim=0cm 0cm 0cm 0cm, clip=false, scale=0.4, angle=0]{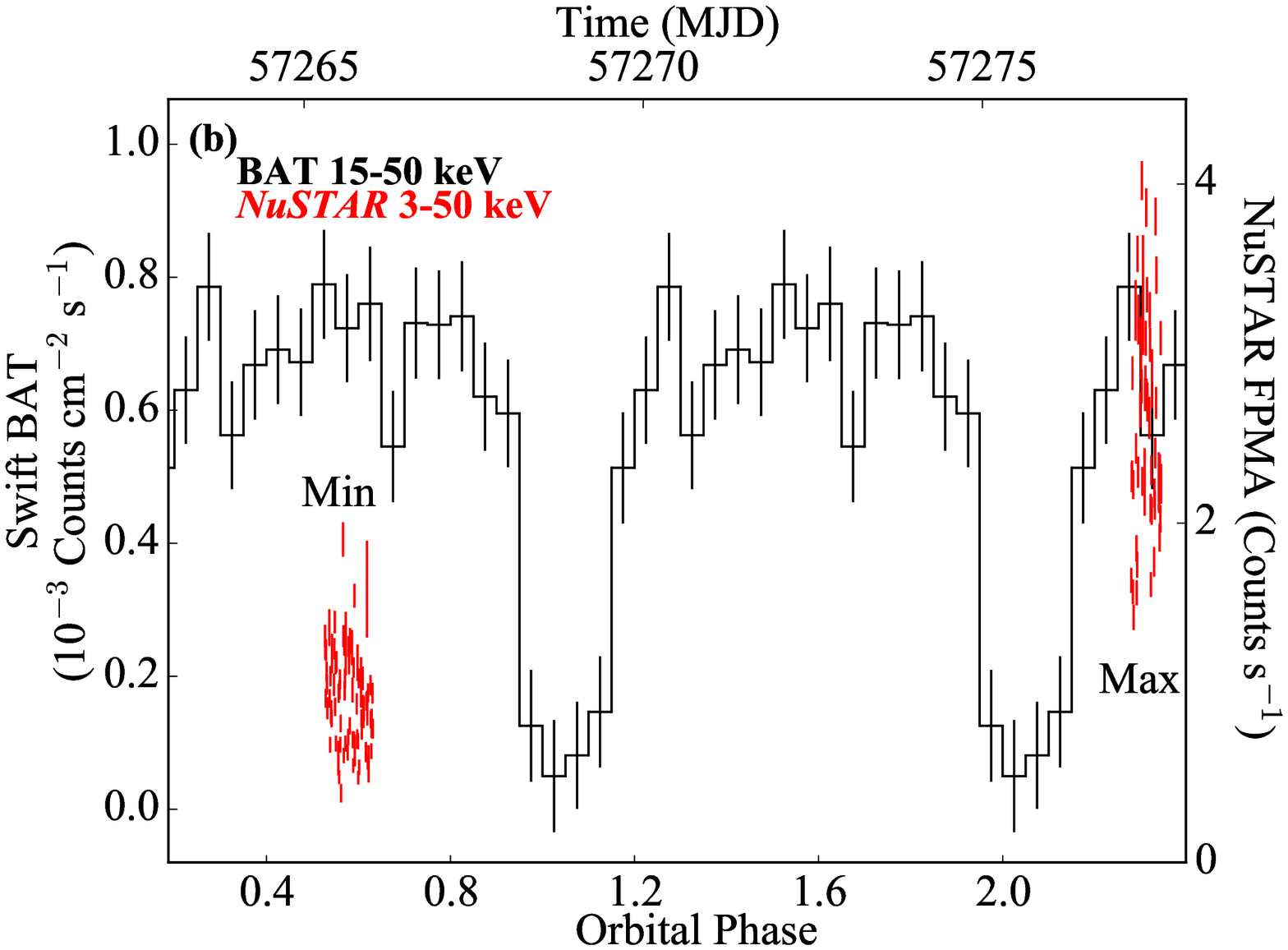}
        \label{Figure:July15-2018_BATNuSTARorbitalprofile.eps}
    }
    \end{tabular}
    \caption{(a) \textsl{Swift} BAT light curve (black) folded on the $\sim$20.06\,day superorbital period using 16 bins.  (b) \textsl{Swift} BAT light curve (black) folded on the $\sim$6.7828\,day orbital period using 20 bins.  Phase zero for the superorbital and orbital light curves corresponds to the times of maximum flux (see Section~\ref{Long-Term Variability}) and maximum delay from the pulsar timing analysis in \citet{2019ApJ...873...86P}, respectively.  The \textsl{NuSTAR} FPMA data near superorbital minimum and maximum are overplotted in red.}
    \label{BAT NuSTAR Overplot}
\end{figure*}

In Figure~\ref{BAT NuSTAR Overplot}(a), we show the \textsl{Swift} BAT light curve folded on the 20.058 $\pm$ 0.007\,day superorbital period using the ephemeris defined in Section~\ref{Long-Term Variability} along with the FPMA light curves, binned to a resolution of {\oldbf 500}\,s.  This illustrates that the \textsl{NuSTAR} observations coincide with superorbital minimum and maximum.

Using the ephemeris defined in \citet{2019ApJ...873...86P}, we also folded the \textsl{Swift} BAT light curve on the 6.7828 $\pm$ 0.0004\,day orbital period (see Figure~\ref{BAT NuSTAR Overplot}{\oldbf (b)}).  The \textsl{NuSTAR} FPMA superorbital minimum and maximum light curves are overplotted in red. This shows that the \textsl{NuSTAR} observations are clearly outside of eclipse.

To determine the neutron star rotation period in both \textsl{NuSTAR} observations, we used the epoch folding technique presented in \citet{1987A&A...180..275L} applied to the combined FPMA$+$FPMB light curves binned to a resolution of 1\,s.  We estimated the uncertainty on the pulse period at the 1$\sigma$ confidence interval by simulating 2000 light curves based on the previously determined pulse period and profile with additional Poisson noise.  We find the neutron star rotation period at superorbital minimum and maximum to be 1092.9 $\pm$ 0.2\,s and 1092.6 $\pm$ 0.3\,s, after correcting for the binary orbital motion.  The neutron star rotation period shows no significant change between superorbital minimum and maximum.  To investigate changes in the neutron star rotation period between our \textsl{NuSTAR} observations and the spin period reported in \citet{2019ApJ...873...86P}, we calculated the Taylor expansion (see Equation~(\ref{Period Derivative Calculation})),

\begin{equation}
\label{Period Derivative Calculation}
P(t)=P(t_0)+(t-t_0) \dot{P}
\end{equation}
where $P(t)$ is the neutron star rotation period derived at superorbital minimum, $P(t_0)$ is the neutron star rotation period derived in \citet{2019ApJ...873...86P}, and the epoch $t_0$ is the maximum delay time from the pulsar timing analysis \citep{2019ApJ...873...86P}.  We find the pulse period derivative between the \textsl{RXTE} observations reported in \citet{2019ApJ...873...86P} and our \textsl{NuSTAR} observation at superorbital minimum to be (-1.8\,$\pm$\,1.7)\,$\times$\,10$^{-9}$\,s s$^{-1}$, which is consistent with zero.

\begin{figure}[h]
\centerline{\includegraphics[width=3in]{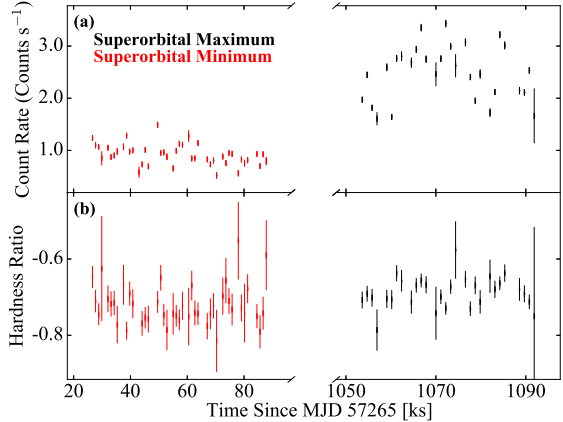}}
\figcaption[April10-2018_FPMAHR.eps]{
(a) \textsl{NuSTAR} FPMA 3--50\,keV light curves and (b) hardness ratio (see text for details) at superorbital minimum (red) and maximum (black) binned to a time resolution corresponding to the average neutron star rotation period ($\sim$1092.7\,s).
\label{NuSTAR Spin Hardness}
}
\end{figure}

In Figure~\ref{NuSTAR Spin Hardness}{\oldbf (a)}, we show the \textsl{NuSTAR} FPMA light curves near superorbital minimum and maximum binned to a time resolution chosen to be the mean pulse period ($\sim$1092.7\,s).  We divided the light curves into two energy bands, where the soft band is defined between energies 3--10\,keV and characterized by the count rate $C_{\rm soft}$, and the hard band is between 15--50\,keV and denoted by the count rate $C_{\rm hard}$.  We define the hardness ratios as:

\begin{equation}
\label{Hardness Equation}
{\rm HR}=\frac{(C_{\rm hard}-C_{\rm soft})}{(C_{\rm hard}+C_{\rm soft})},
\end{equation}
where a soft spectrum is indicated by negative values and a hard spectrum is indicated by positive values (see Figure~\ref{NuSTAR Spin Hardness}{\oldbf (b)}).  While the X-ray flux significantly increases between superorbital minimum and maximum, no change in the hardness ratio between superorbital minimum and maximum was found.

\begin{figure}[h]
\centerline{\includegraphics[width=3in]{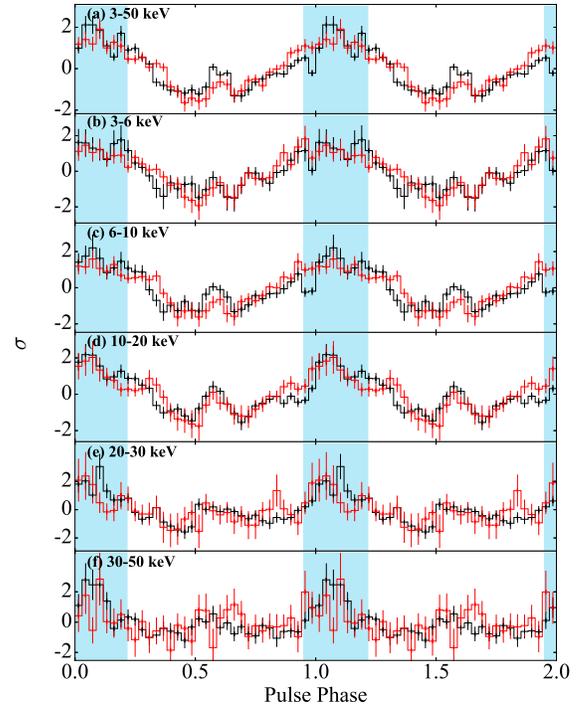}}
\figcaption[April25-2018_NuSTARfullprofileb.eps]{
Energy-resolved \textsl{NuSTAR} pulse profiles at superorbital minimum (red) and superorbital maximum (black).  The pulse profiles are normalized such that their mean value is zero and their standard deviation is unity.  The modulation appears to be double-peaked, with a main broad peak and secondary peak offset by $\sim$0.5 in phase.  The phase range used for the pulse-peak spectral analysis is indicated by the light blue shaded region (see Section~\ref{Pulse-peak Spectral Analysis}).
\label{Energy Resolved Spin Profiles}
}
\end{figure}

We folded the \textsl{NuSTAR} 3--50\,keV light curves at superorbital minimum and maximum on the $\sim$1092.9\,s and $\sim$1092.6\,s periods, respectively. For the observation at superorbital maximum, we defined phase zero at the time of maximum delay (see Section~\ref{Introduction}).  We aligned the pulse profiles at superorbital minimum and maximum by calculating the maximum value of the cross correlation function between the two pulse profiles (see Figure~\ref{Energy Resolved Spin Profiles}).  The pulse profiles at superorbital minimum and maximum each show a double-peaked structure with a main broad peak and a smaller secondary peak.

To investigate the energy dependence of the pulse profile, we divided the light curve into five energy bands defined between energies of 3--6\,keV, 6--10\,keV, 10--20\,keV, 20--30\,keV and 30--50\,keV, respectively.  The pulse profiles at both superorbital minimum and maximum show a weak energy dependence (see Figure~\ref{Energy Resolved Spin Profiles}).  Only small changes in the pulse profiles are seen, where the main peak is broad up to 20\,keV and progressively {\oldbf becomes narrower} up to $\sim$50\,keV (see Figure~\ref{Energy Resolved Spin Profiles}{\oldbf (b)}--{\oldbf (f)}).

\begin{figure}[ht]
\centerline{\includegraphics[width=3in]{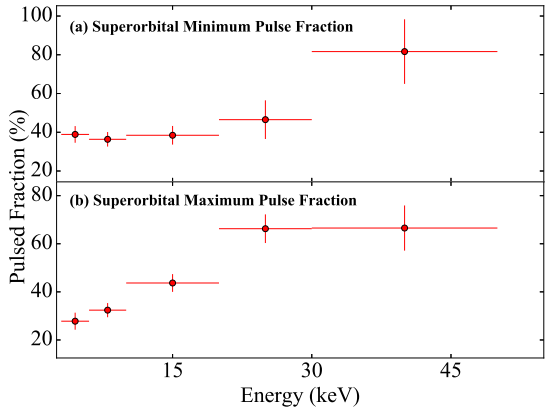}}
\figcaption[April10-2018_NuSTARPFsuper.eps]{
Energy dependence of the pulse fraction of IGR J16493-4348 observed near (a) superorbital minimum and (b) superorbital maximum.
\label{NuSTAR Pulse Fraction}
}
\end{figure}

We define the peak-to-peak pulse fraction as:

\begin{equation}
\label{Pulse Fraction Equation}
\mathcal{P}=\frac{(F_{\rm max}-F_{\rm min})}{(F_{\rm max}+F_{\rm min})}
\end{equation}
where the maximum and minimum count rates in the pulse profile are characterized as $F_{\rm max}$ and $F_{\rm min}$, respectively.  Using Equation~(\ref{Pulse Fraction Equation}), we found that the peak-to-peak pulsed fraction at both superorbital minimum and maximum increases with increasing energy (see Figure~\ref{NuSTAR Pulse Fraction}).

\subsection{Phase-Averaged Spectral Analysis}
\label{Phase-Averaged Spectral Analysis}

The nearly simultaneous \textsl{NuSTAR} and \textsl{Swift} XRT spectra of IGR J16493-4348 were analyzed using the package \texttt{XSPEC v12.9.1}.  We made use of the \texttt{XSPEC} convolution model \texttt{cflux} to calculate the fluxes and associated errors.  The FPMA and FPMB spectra were simultaneously fit in XSPEC.  To account for instrumental calibration uncertainties, we used cross-calibration constants normalized to FPMA during the spectral analysis (see Table~\ref{Broadband Spectral Parameters}).

\begin{figure*}[h]
    \centering
    \begin{tabular}{cc}
    \subfigure
    {
        \includegraphics[trim=0cm 0cm 0cm 0cm, clip=false, scale=0.3, angle=0]{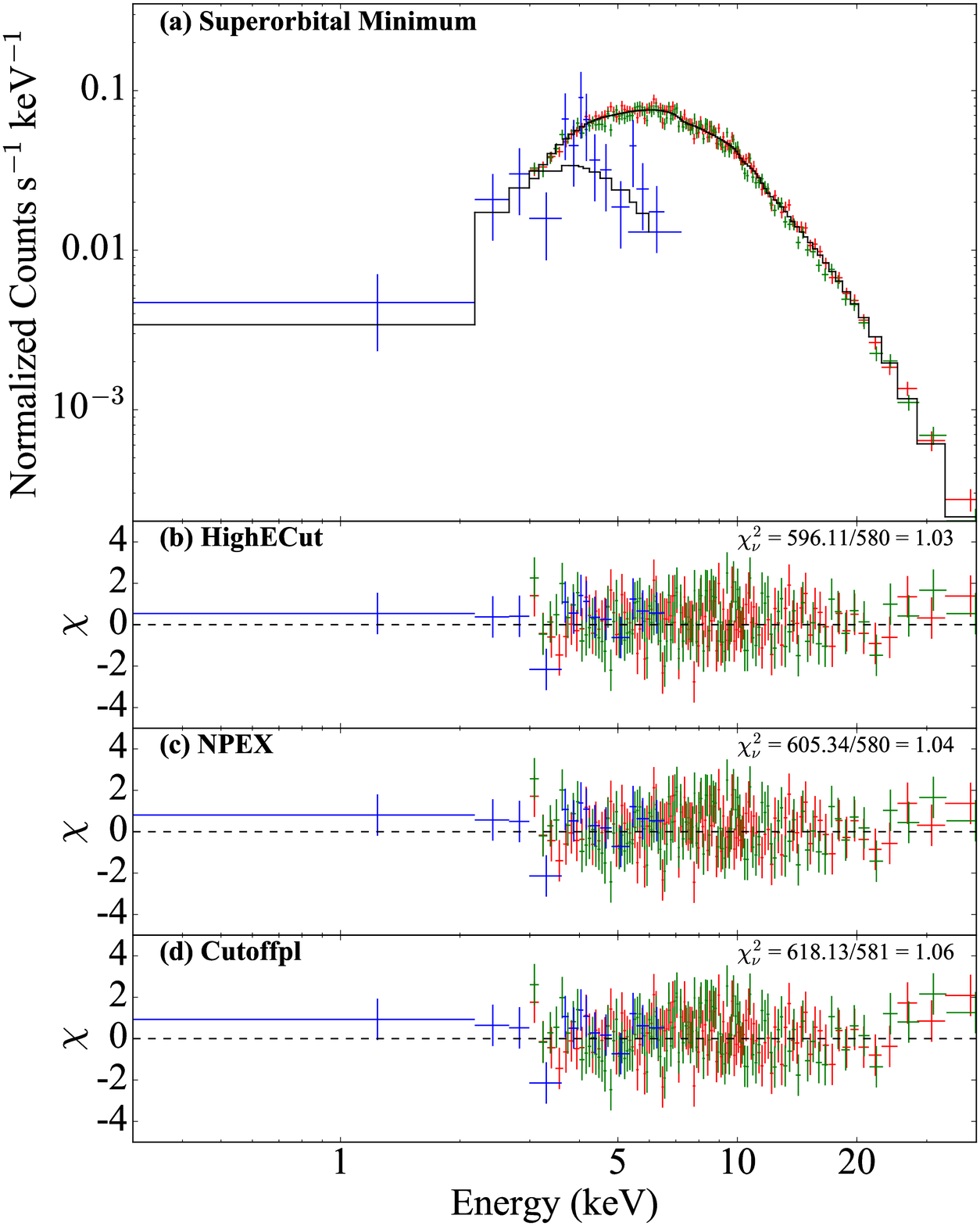}
        \label{Figure:4-12-2018_ObsIRebinSpectrum.eps}
    }
    &
    \subfigure
    {
        \includegraphics[trim=0cm 0cm 0cm 0cm, clip=false, scale=0.3, angle=0]{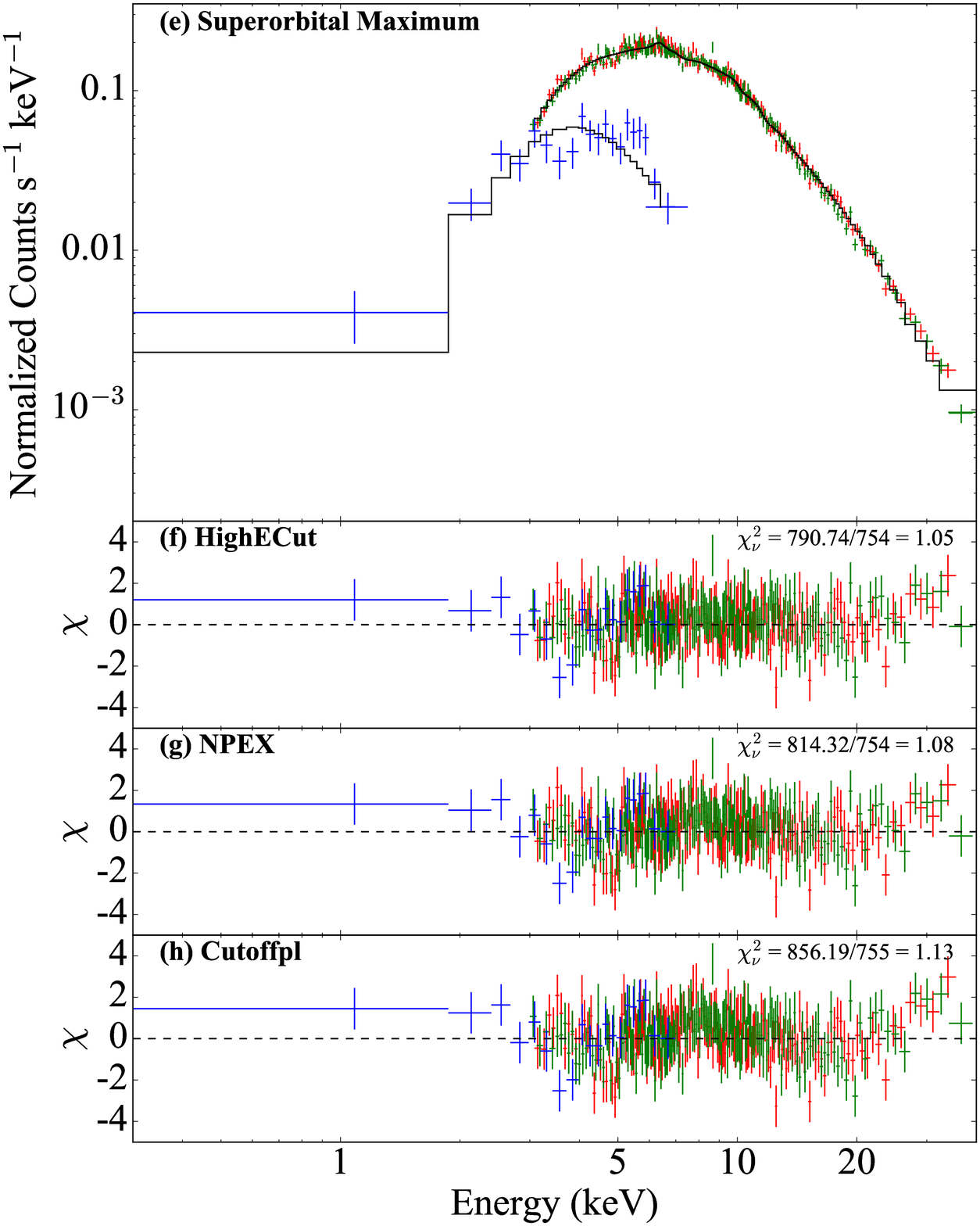}
        \label{Figure:4-12-2018_ObsIIRebinSpectrum.eps}
    }
    \end{tabular}
    \caption{Broadband \textsl{NuSTAR}$+$\textsl{Swift} spectra of IGR J16493-4348 at (a) superorbital minimum and (e) superorbital maximum where the FPMA, FPMB, and XRT data are shown in red, green, and blue, respectively.  The best fit \texttt{highecut} model is shown in black.  Both models consist of a continuum comprised of an absorbed power law with a high-energy cutoff and an emission line near 6.4\,keV.  Residuals of the best fit \texttt{highecut} model are plotted for (b) superorbital minimum and (f) superorbital maximum, respectively.  Residuals for the \texttt{NPEX} model are plotted for (c) superorbital minimum and (g) superorbital maximum, respectively.  Residuals for the \texttt{cutoffpl} model are plotted for (d) superorbital minimum and (h) superorbital maximum, respectively.  {\mybf The spectra are rebinned for clarity.}}
    \label{NuSTAR Swift Broadband Spectra}
\end{figure*}

To monitor the spectral evolution of the broadband X-ray emission as a function of superorbital phase, we extracted nearly simultaneous \textsl{NuSTAR} and \textsl{Swift} XRT spectra at superorbital minimum and maximum, respectively (see Figure~\ref{NuSTAR Swift Broadband Spectra}(a) and (e)).  For both datasets, we used several models that typically describe systems that host a neutron star: a power law (\texttt{power}), a power law with a high-energy cutoff \citep[\texttt{highecut};][]{1983ApJ...270..711W}, a cutoff power-law (\texttt{cutoffpl}), a power law with a Fermi-Dirac cutoff \citep[\texttt{fdcut};][]{1986LNP...255..198T}, and a negative-positive exponential cutoff \citep[\texttt{npex};][]{1999ApJ...525..978M}.  All models were modified by an absorber that fully covers the source (\texttt{tbabs} in XSPEC) using the \citet{1996ApJ...465..487V} cross sections and \citet{2000ApJ...542..914W} abundances.  We note \citet{DAi2011} applied the \texttt{npex}, \texttt{cutoffpl} and \texttt{fdcut} models to a broadband analysis using \textsl{Swift} BAT and \textsl{INTEGRAL} ISGRI data, together with pointed \textsl{Swift} XRT and \textsl{Suzaku} observations, and found the spectra to be best described by the \texttt{npex} model.  For our \textsl{NuSTAR} and \textsl{Swift} observations, we could not constrain the cutoff energy with the \texttt{fdcut} model and find an upper limit at the 90$\%$ confidence interval of $<$3\,keV at both superorbital minimum and maximum.

\begin{deluxetable*}{ccccccc}
\tablecolumns{7}
\tablewidth{0pc}
\tablecaption{Phase-averaged, broadband X-ray spectral parameters of the nearly simultaneous \textsl{NuSTAR} and \textsl{Swift} Observations for several empirical models}
\tablehead{
\colhead{Model Parameter} & \multicolumn{2}{c}{\texttt{Highecut}} & \multicolumn{2}{c}{\texttt{NPEX}} & \multicolumn{2}{c}{\texttt{CutoffPL}} \\
\colhead{} & \colhead{Superorbital} & \colhead{Superorbital} & \colhead{Superorbital} & \colhead{Superorbital} & \colhead{Superorbital} & \colhead{Superorbital} \\
\colhead{} & \colhead{Minimum} & \colhead{Maximum} & \colhead{Minimum} & \colhead{Maximum} & \colhead{Minimum} & \colhead{Maximum}} \\
\startdata
$\chi_\nu^2$ (dof) & 1.03 (581) & 1.05 (753) & 1.04 (580) & 1.08 (753) & 1.06 (581) & 1.13 (754) \\
$C_{\rm FPMA}$$^a$ & 1 & 1 & 1 & 1 & 1 & 1 \\
$C_{\rm FPMB}$$^a$ & 1.03 $\pm$ 0.02 & 1.03 $\pm$ 0.01 & 1.03 $\pm$ 0.02 & 1.03 $\pm$ 0.01 & 1.03 $\pm$ 0.02 & 1.03 $\pm$ 0.01 \\
$C_{\rm XRT}$$^a$ & 0.9 $\pm$ 0.2 & 0.80 $\pm$ 0.07 & 0.9 $\pm$ 0.2 & 0.80 $\pm$ 0.07 & 0.9 $\pm$ 0.2 & 0.80 $\pm$ 0.07 \\
Cutoff Energy (keV) & 6.9$^{+1.3}_{-0.7}$ & 8.1$^{+0.4}_{-0.5}$ & \nodata & \nodata & \nodata & \nodata \\
Folding Energy (keV) & 15$^{+3}_{-2}$ & 19$\pm$2 & 8$^{+2}_{-1}$ & 8.7$^{+1.1}_{-0.9}$ & 12$^{+2}_{-1}$ & 14 $\pm$ 1 \\
Norm$_{\rm n}$ ($\times$10$^{-4}$) & \nodata & \nodata & 0.9$^{+1.6}_{-0.7}$ & 0.9$^{+0.7}_{-0.5}$ & \nodata & \nodata \\
Tbabs $N_{\rm H}$ ($\times$10$^{22}$\,cm$^{-2}$)& 9 $\pm$ 2 & 11 $\pm$ 1 & 9$^{+2}_{-1}$ & 10 $\pm$ 1 & 10 $\pm$ 2 & 11 $\pm$ 1 \\
$\Gamma$ & 1.3 $\pm$ 0.2 & 1.31$^{+0.07}_{-0.08}$ & 0.7$^{+0.2}_{-0.1}$ & 0.72$^{+0.10}_{-0.09}$ & 1.0 $\pm$ 0.1 & 0.99 $\pm$ 0.09 \\
Normalization ($\times$10$^{-2}$) & 0.36$^{+0.17}_{-0.09}$ & 0.9$^{+0.2}_{-0.1}$ &  0.29$^{+0.07}_{-0.05}$ & 0.65$^{+0.09}_{-0.08}$ & 0.39$^{+0.09}_{-0.07}$ & 0.8 $\pm$ 0.1 \\
\tableline
Fe K$\alpha$ Energy (keV) & 6.36$^b$ & 6.36$^{+0.09}_{-0.10}$ & 6.4$^b$ & 6.4 $\pm$ 0.1 & 6.4$^b$ & 6.4 $\pm$ 0.1 \\
Fe K$\alpha$ Width ($\sigma_{\rm Fe K\alpha}$)$^c$ & 0.1 & 0.1 & 0.1 & 0.1 & 0.1 & 0.1 \\
Normalization ($\times$10$^{-3}$\,photons cm$^{-2}$ s$^{-1}$) & $<$0.01 & 0.04 $\pm$ 0.02 & $<$0.01 & 0.04 $\pm$ 0.02 & $<$0.01 & 0.04 $\pm$ 0.02 \\
Fe K$\alpha$ EQW (eV) & $<$44 & 51$^{+22}_{-19}$ & $<$44 & 44$^{+32}_{-18}$ & $<$44 & 47$^{+21}_{-19}$ \\
Fe K$\alpha$ Flux ($\times$10$^{-13}$\,erg cm$^{-2}$ s$^{-1}$) & $<$0.8 & 3$^{+1}_{-2}$ & $<$1.0 & 2.5$^{+0.9}_{-1.8}$ & $<$0.7 & 3$^{+1}_{-2.9}$ \\
\tableline
Absorbed Flux ($\times$10$^{-11}$\,erg cm$^{-2}$ s$^{-1}$)$^d$ & 1.98 $\pm$ 0.03 & 4.90 $\pm$ 0.06 & 1.96 $\pm$ 0.03 & 4.87 $\pm$ 0.05 & 1.95 $\pm$ 0.03 & 4.85 $\pm$ 0.05 \\
Unabsorbed Flux ($\times$10$^{-11}$\,erg cm$^{-2}$ s$^{-1}$)$^e$ & 2.9 $\pm$ 0.3 & 7.6 $\pm$ 0.2 & 2.9$^{+0.2}_{-0.1}$ & 7.3 $\pm$ 0.3 & 3.1 $\pm$ 0.2 & 7.7 $\pm$ 0.3 \\
\enddata
\tablecomments{\\*
$^a$ Detector cross-calibration constants with respect to FPMA. \\*
$^b$ The energy is frozen because we can only obtain an upper limit. \\*
$^c$ The width of the Fe K$\alpha$ line is frozen to 0.1\,keV. \\*
$^d$ Absorbed flux in the 1--10\,keV band. \\*
$^e$ Unabsorbed flux in the 1--10\,keV band.}
\label{Broadband Spectral Parameters}
\end{deluxetable*}

We compared the \texttt{highecut}, \texttt{npex}, and \texttt{cutoffpl} residuals at superorbital minimum and maximum in Figures~\ref{NuSTAR Swift Broadband Spectra}(b)--(d) and~\ref{NuSTAR Swift Broadband Spectra}(f)--(h), respectively, and found that all three models provided a similar quality of fit at superorbital minimum (see Figure~\ref{NuSTAR Swift Broadband Spectra}a).  The \texttt{highecut} and \texttt{npex} models provided a similar quality of fit at superorbital maximum, but the \texttt{cutoffpl} model yielded a worse $\chi^2$ value and wavy residuals (see Figure~\ref{NuSTAR Swift Broadband Spectra}{\oldbf (h)}).  The results are given in Table~\ref{Broadband Spectral Parameters}.  We note that for a power law modified by a high-energy cutoff, sharp features can appear as line-like residuals at the cutoff energy.  No evidence of {\oldbf such} line-like residuals near the cutoff energy was found in either observation (see Figure~\ref{NuSTAR Swift Broadband Spectra}).  Unless otherwise noted, we chose to model the spectra using the \texttt{highecut} model since it provided a marginally better fit quality at both superorbital minimum and maximum (see Figures~\ref{NuSTAR Swift Broadband Spectra}{\oldbf (b)} and~\ref{NuSTAR Swift Broadband Spectra}{\oldbf (f)}).

The neutral hydrogen column density for the fully covered absorption at superorbital minimum and maximum were found to be (9 $\pm$ 2)$\times$10$^{22}$\,cm$^{-2}$ and (11 $\pm$ 1)$\times$10$^{22}$\,cm$^{-2}$, respectively.  These measurements exceed the values reported by the Leiden/Argentine/Bonn survey \citep{2005A&A...440..775K} and the review by \citet{1990ARA&A..28..215D}, which are 1.42\,$\times$\,10$^{22}$\,cm$^{-2}$ and 1.82\,$\times$\,10$^{22}$\,cm$^{-2}$, respectively.  This is consistent with absorbing material intrinsic to the source, which is expected for the subclass of obscured SGXBs \citep{2011ASPC..447...29C}.

Apart from the overall flux change, we find no significant changes in the continuum parameters between superorbital minimum and maximum (see Table~\ref{Broadband Spectral Parameters}).  Not surprisingly, the unabsorbed X-ray flux is found to increase by more than a factor of 2, where it is found to be (2.9 $\pm$ 0.3)\,$\times$\,10$^{-11}$\,erg cm$^{-2}$ s$^{-1}$ in the 1--10\,keV band near superorbital minimum and (7.6 $\pm$ 0.2)\,$\times$\,10$^{-11}$\,erg cm$^{-2}$ s$^{-1}$ in the 1--10\,keV band near superorbital maximum.  Assuming a distance of {\oldbf 16.1\,$\pm$\,1.5}\,kpc \citep{2019ApJ...873...86P}, the 1--10\,keV X-ray luminosity is found to be {\oldbf (9\,$\pm$\,2)\,$\times$\,10$^{35}$\,erg s$^{-1}$ and (2.4\,$\pm$\,0.5)\,$\times$\,10$^{36}$\,erg s$^{-1}$ at superorbital minimum and maximum, respectively.}

{\oldbf To study the broadband behavior of IGR J16493-4348, we additionally calculated the unabsorbed X-ray flux and luminosity in the 3--40\,keV band.  At superorbital minimum, the 3--40\,keV X-ray flux and luminosity are found to be (4.9 $\pm$ 0.2)\,$\times$\,10$^{-11}$\,erg cm$^{-2}$ s$^{-1}$ and (1.5\,$\pm$\,0.3)\,$\times$\,10$^{36}$\,erg s$^{-1}$, respectively.  The corresponding X-ray flux and luminosity at superorbital maximum are ({\mybf 1.41\,$\pm$\,0.03})\,$\times$\,10$^{-10}$\,erg cm$^{-2}$ s$^{-1}$ and (4.4\,$\pm$\,0.8)\,$\times$\,10$^{36}$\,erg s$^{-1}$, respectively.}

\begin{figure}[h]
\centerline{\includegraphics[width=3in]{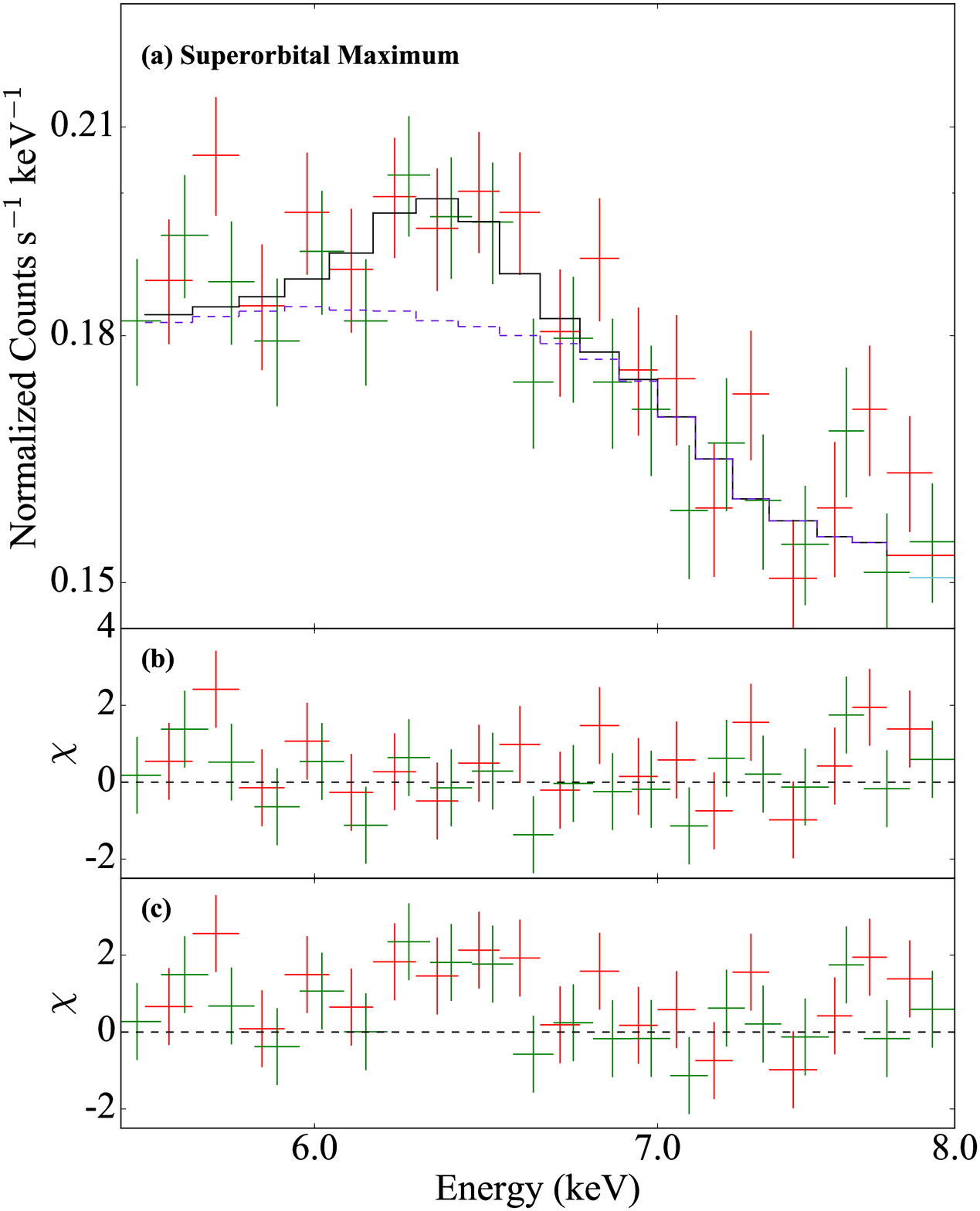}}
\figcaption[April12-2018_Spectrum.eps]{
The FPMA$+$FPMB spectrum of IGR J16493-4348 between 5.5-8.0\,keV band at superorbital maximum.  The Fe K$\alpha$ emission line is shown along with the best fit model (black).  (a) FPMA and FPMB data are indicated by the red and green data points. The dashed purple line indicates the fit without the Fe K$\alpha$ emission line.  (b) Residuals with the Fe K$\alpha$ emission line included in the spectral model. (c) Residuals without the Fe K$\alpha$ emission line included in the spectral model.
\label{Fe K zoom}
}
\end{figure}

Some residuals were found near 6.4\,keV at superorbital maximum, which could indicate a weak Fe K$\alpha$ {\mybf emission} feature (see Figure~\ref{Fe K zoom}). {\mybf We account for this with a narrow additive Gaussian with a line width fixed to 0.1\,keV, as the line width was unconstrained by the fit.  We also tried freezing the width of the Fe K$\alpha$ line to 0.01\,keV and 10$^{-3}$\,keV, but note that it did not significantly affect the best fit continuum parameters and their uncertainties.  This is not surprising since these line widths are smaller than \textsl{NuSTAR}'s FWHM energy resolution, which is 400\,eV at 6.0\,keV \citep{2013ApJ...770..103H}.}

The addition of an Fe K$\alpha$ line reduces the $\chi^2/$d.o.f. from 807.27/755 to 791.14/753.  To estimate the significance of the inclusion of an Fe K$\alpha$ feature to the \textsl{NuSTAR} and \textsl{Swift} spectra, we simulated 10$^{4}$ spectra using a Monte Carlo analysis \citep{2002ApJ...571..545P}.  The simulated spectra were modeled without the Fe $K\alpha$ emission component and were fit both with and without the additional component.  We compared the difference in simulated $\chi^{2}$ with the observed one, which was found to be 16.1.  We find the significance of an Fe K$\alpha$ feature to be 99.99$\%$, which supports the presence of a neutral Fe $K\alpha$ feature.  We find its centroid energy, equivalent width, and flux are 6.36$^{+0.09}_{-0.10}$\,keV, 51$^{+22}_{-19}$\,eV, and (3$^{+1}_{-2}$)$\times$10$^{-13}$\,erg cm$^{-2}$ s$^{-1}$, respectively (see Table~\ref{Broadband Spectral Parameters}).

At superorbital minimum, we find the addition of an Fe K$\alpha$ feature does not significantly improve the fit quality ($\chi^2/$d.o.f. changes from 596.17/581 to 596.12/580).  Using {\oldbf a similar Monte Carlo analysis}, we determined the significance of an Fe K$\alpha$ feature to be less than 68$\%$ with 10$^{4}$ trials.  From a spectral analysis using \textsl{Suzaku} and \textsl{Swift} BAT observations, \citet{2009ApJ...699..892M} found no evidence of strong Fe K$\alpha$ features and calculated the upper limit of the equivalent width of a 6.4\,keV line to be 84\,eV.  To fit the spectrum with the same model in both observations, we chose to include the Fe K$\alpha$ line where the energy and width were fixed to the best fit values found at superorbital maximum.  We find an upper limit of 44\,eV and 8.0$\times$10$^{-14}$\,erg cm$^{-2}$ s$^{-1}$ {\oldbf at the 90$\%$ confidence interval} for the equivalent width and flux of the Fe K$\alpha$ line, respectively (see Table~\ref{Broadband Spectral Parameters}).

Negative residuals were found near $\sim$20\,keV at superorbital minimum, indicating the possible presence of a narrow absorption feature (see Figure~\ref{NuSTAR Swift Broadband Spectra}{\oldbf (b)}).  We modeled the residuals near $\sim$20\,keV using a multiplicative line model with the centroid energy, line width, and optical depth as free parameters.  We investigated the significance of a $\sim$20 keV feature using the Monte Carlo analysis described in \citet{2002ApJ...571..545P} and determined its significance to be less than $\sim$60$\%$ with 10$^{4}$ trials, which indicates the improvement from adding an absorption component near $\sim$20\,keV is negligible.  We do not find broad residuals between 30\,keV and 40\,keV in the \textsl{NuSTAR} spectra, even though such residuals were significant in the BAT and ISGRI spectra reported in \citet{DAi2011}.

\subsection{Pulse-Phase Resolved Spectral Analysis}
\label{Phase-Resolved Spectral Analysis}

We investigated variations in the spectral continuum at different rotational phases of the neutron star using pulse-phase resolved spectroscopy.  Since the exposure times of the \textsl{Swift} snapshot observations were short in comparison to the \textsl{NuSTAR} observations (see Table~\ref{X-ray Observations Summary}), we only considered the \textsl{NuSTAR} spectra for the analysis of phase dependent changes in the spectral parameters.  For the phase-resolved analysis, we chose to subdivide the folded light curves at superorbital minimum and maximum into four equally spaced intervals.  We rebinned the phase-resolved spectra using the same procedure as for the phase-averaged spectra (see Section~\ref{NuSTAR description}).

\begin{figure}[h]
\centerline{\includegraphics[width=3.3in]{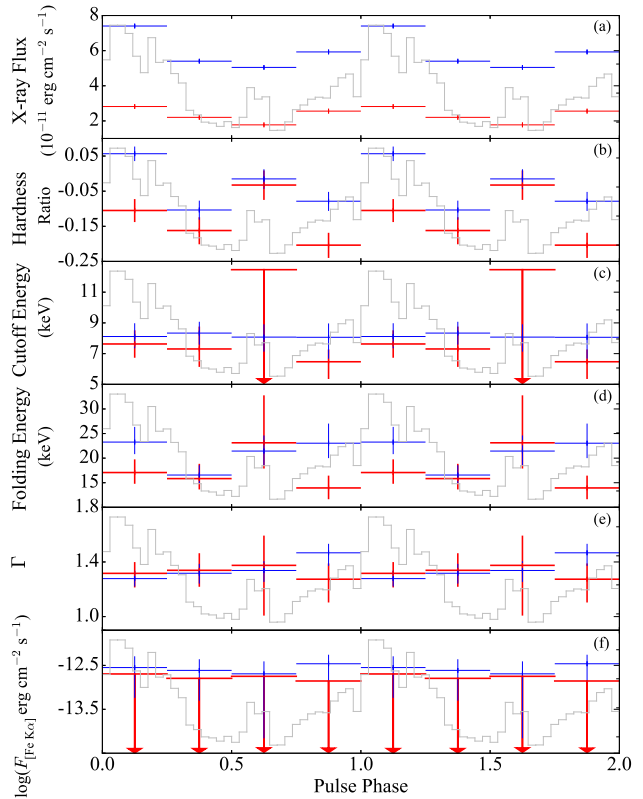}}
\figcaption[May102018_phaseresolvedb.eps]{
\textsl{NuSTAR} spectral parameters as a function of pulse phase using the power law with high-energy cutoff model.  The observations at superorbital minimum and maximum are indicated by the red and blue points, respectively.  The grey histogram shows the \textsl{NuSTAR} light curve at superorbital maximum in the 3--50\,keV energy range folded on the neutron star rotation period.
\label{Pulse Resolved Max Spectral Fits}
}
\end{figure}

We initially allowed the continuum spectral components described in Table~\ref{Broadband Spectral Parameters} to be free parameters and performed spectral fits on each of the four equally spaced intervals.  Since the phase-resolved spectra lack the soft energy coverage made available by \textsl{Swift} XRT, we chose to fix $N_{\rm H}$ to the phase-averaged values of 9$\times$10$^{22}$\,cm$^{-2}$ and 11$\times$10$^{22}$\,cm$^{-2}$ from the superorbital minimum and maximum spectra, respectively (see Table~\ref{Broadband Spectral Parameters}).  We also tried to fit the data leaving the fully covered $N_{\rm H}$ free, but this resulted in large uncertainties in the model parameters.

Figure~\ref{Pulse Resolved Max Spectral Fits} shows the spectral parameters of the \texttt{highecut} model at different rotational phases of the neutron star.  {\oldbf We find} possible evidence of an increase in the folding energy between superorbital minimum and maximum near the main peak of the pulse profile (see Figure~\ref{Pulse Resolved Max Spectral Fits}(d)).

\begin{figure}[h]
\centerline{\includegraphics[width=3in]{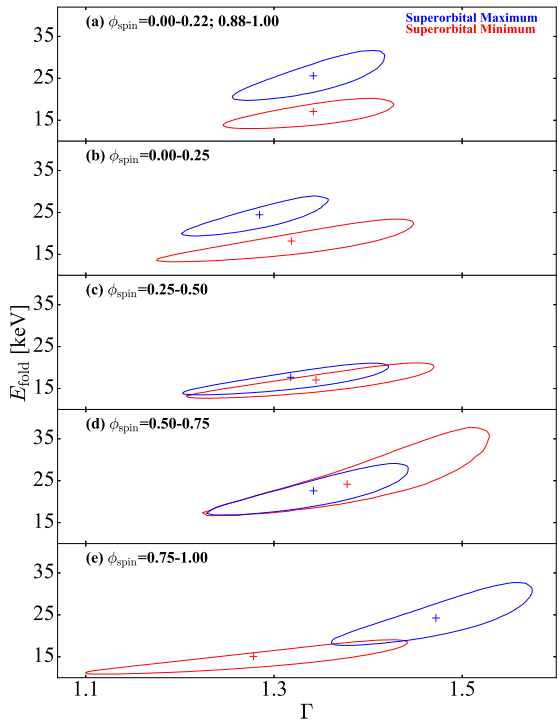}}
\figcaption[July182018_SuperorbitalGammaEfold.eps]{
Contours of the folding energy, $E_{\rm fold}$, and photon index, $\Gamma$, at the 3$\sigma$ confidence interval for the phase-resolved spectra at superorbital minimum {\oldbf (red)} and maximum {\oldbf (blue)}.  The neutral hydrogen column density, Fe K$\alpha$ energy, and detector calibration constants were held constant.  {\oldbf The best fit values at superorbital minimum and maximum are indicated by the red and blue crosses, respectively.}
\label{Contour Plots}
}
\end{figure}

In Figure~\ref{Contour Plots}, we show the 3$\sigma$ confidence contours between the folding energy and the photon index at both superorbital minimum and maximum {\oldbf at each neutron star rotation period phase bin}.  The folding energy and photon index are found to be roughly correlated with each other.  {\oldbf We found the folding energy significantly increases between superorbital minimum and maximum near the main peak of the pulse profile (see Figure~\ref{Contour Plots}(a)--(b)),} confirming our results in Figure~\ref{Pulse Resolved Max Spectral Fits}{\oldbf (d)}.

To investigate possible changes in the {\oldbf pulse-phase resolved} spectral shape between superorbital minimum and maximum, we calculated the hardness ratio using Equation~(\ref{Hardness Equation}).  We defined the soft and hard bands to be between 3--10\,keV and 15--40\,keV, respectively.  The hardness ratio at the main peak of the pulse profile increases with increasing X-ray luminosity (see Figure~\ref{Pulse Resolved Max Spectral Fits}{\oldbf (b)}).

Due to the reduced signal-to-noise compared to the phase-averaged spectrum, the addition of a 6.4\,keV emission feature does not significantly improve the quality of the fit in most phase intervals of the the phase-resolved spectra at superorbital maximum, even though it was observed at the 99.99$\%$ confidence interval in the superorbital maximum phase-averaged spectrum (see Section~\ref{Phase-Averaged Spectral Analysis}).  We note the Fe K$\alpha$ line is significant at the 98.8$\%$ confidence intervals between pulse phases 0.75--1.00 using a Monte Carlo analysis with 10$^{4}$ trials.  Since the Fe K$\alpha$ line is detected in the phase-averaged spectrum at superorbital maximum, we chose to include it in our {\mybf pulse-}phase-resolved spectra at both superorbital maximum and superorbital minimum with the centroid energy and width frozen to the value determined from the phase-averaged spectrum at superorbital maximum.  No fluctuations in the flux of the Fe K$\alpha$ line as a function of neutron star rotation period were found (see Figure~\ref{Pulse Resolved Max Spectral Fits}{\oldbf (f)}), which could possibly be attributed to the low signal to noise.

\subsection{Pulse-peak Spectral Analysis}
\label{Pulse-peak Spectral Analysis}

\begin{figure}[h]
\centerline{\includegraphics[width=3.in]{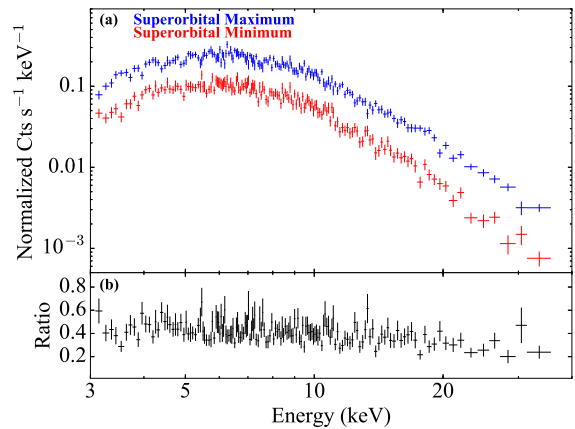}}
\figcaption[May16-2018_PulsePeakFPMARatiobin40.eps]{
(a) Pulse-peak FPMA spectra at superorbital minimum (red) and superorbital maximum (blue).  The superorbital maximum spectrum was rebinned to match the energy binning of the superorbital minimum spectrum for the plot to compare the residuals on a bin-by-bin basis.  (b) Count Rate spectral ratio between superorbital minimum and maximum (see text for details).
\label{Pulse Peak Ratio}
}
\end{figure}

To further investigate possible spectral differences between superorbital minimum and maximum near the broad main peak of the pulse profile, we extracted a spectrum for each observation focusing on pulse phases 0.00--0.22 and 0.88--1.00.  We rebinned the pulse-peak resolved spectra using the same procedure as for the phase-averaged and phase-resolved spectra and again only considered the \textsl{NuSTAR} spectra for the analysis.

For the pulse-peak spectral analysis, we fit the spectra at both superorbital minimum and maximum simultaneously using the procedure described in \citet[][and references therein]{2016AcPol..56...41K}.  To reduce the number of free parameters, we froze the energy and width of the Fe K$\alpha$ feature to the phase-averaged value at superorbital maximum.  Again, the Fe K$\alpha$ feature was not detected at superorbital minimum and we therefore derived an upper limit for the strength of the line.

In Figure~\ref{Pulse Peak Ratio}, we show the pulse-peak \textsl{NuSTAR} FPMA spectra and count rate ratio at superorbital minimum and maximum.  At energies above 10\,keV, we find possible evidence of a harder spectrum at superorbital maximum than superorbital minimum (see Figure~\ref{Pulse Peak Ratio}(b)).  This is consistent with the increase in the hardness ratio from superorbital minimum to superorbital maximum near the peak of the pulse profile (see Figure~\ref{Pulse Resolved Max Spectral Fits}(b)).  To investigate changes in the shape of the spectrum between superorbital minimum and maximum, we first fit the pulse-peak spectra at both superorbital phases simultaneously and only allowed the cross-normalization for each spectrum to change.  In these fits, the continuum parameters were all tied together to have the same value for the superorbital minimum and maximum spectra \cite[see][and references therein]{2018ApJ...852..132B}.  We find $\chi^2_{\nu}$ to be 1.14 for 838 d.o.f., 1.15 for 838 d.o.f. and 1.18 for 837 d.o.f. for the \texttt{highecut}, \texttt{npex}, and \texttt{cutoffpl} models, respectively.  The cutoff energy, folding energy and photon index for the \texttt{highecut} model were found to be 8.0\,$\pm$\,0.3\,keV, 22\,$\pm$\,1\,keV, and 1.35\,$\pm$\,0.03, respectively.

We also fit {\oldbf the} pulse-peak spectra at superorbital minimum and {\oldbf maximum}, where the folding energy was allowed to vary.  This reduces the $\chi^2_{\nu}$ for the \texttt{highecut}, \texttt{npex}, and \texttt{cutoffpl} models {\oldbf to 1.01 for 836 d.o.f., 1.02 for 836 d.o.f. and 1.05 for {\mybf 835} d.o.f., respectively (see Table~\ref{Pulse Peak Spectral Parameters}).}  The folding energy for the \texttt{highecut} and \texttt{cutoffpl} models shows a possible increase between superorbital minimum and maximum, which is consistent with our pulse-phase resolved results (see Figure~\ref{Pulse Resolved Max Spectral Fits}(b)).

\begin{deluxetable*}{ccccccc}
\tablecolumns{7}
\tablewidth{0pc}
\tablecaption{Pulse-Peak Resolved Broadband X-ray Spectral Parameters}
\tablehead{
\colhead{Model Parameter} & \multicolumn{2}{c}{\texttt{Highecut}} & \multicolumn{2}{c}{\texttt{NPEX}} & \multicolumn{2}{c}{\texttt{CutoffPL}} \\
\colhead{} & \colhead{Superorbital} & \colhead{Superorbital} & \colhead{Superorbital} & \colhead{Superorbital} & \colhead{Superorbital} & \colhead{Superorbital} \\
\colhead{} & \colhead{Minimum} & \colhead{Maximum} & \colhead{Minimum} & \colhead{Maximum} & \colhead{Minimum} & \colhead{Maximum}} \\
\startdata
$\chi_\nu^2$ (dof) & \multicolumn{2}{c}{1.01 (836)} & \multicolumn{2}{c}{1.02 (836)} & \multicolumn{2}{c}{1.05 ({\mybf 835})} \\
$C_{\rm FPMA}$ & 1$^a$ & {\oldbf 2.44 $\pm$ 0.06}$^a$ & 1$^a$ & {\oldbf 2.12 $\pm$ 0.08}$^a$ & 1$^a$ & {\oldbf 2.13 $\pm$ 0.09}$^a$ \\
$C_{\rm FPMB}$ & 1.03 $\pm$ 0.02$^a$ & {\oldbf 2.51 $\pm$ 0.06}$^a$ & 1.03 $\pm$ 0.02$^a$ & {\oldbf 2.18$^{+0.09}_{-0.08}$$^a$} & {\oldbf 1.03 $\pm$ 0.02}$^a$ & {\oldbf 2.19 $\pm$ 0.09}$^a$ \\
Cutoff Energy (keV) & \multicolumn{2}{c}{\oldbf 7.7$^{+0.7}_{-0.9}$} & \nodata & \nodata & \nodata & \nodata \\
Folding Energy (keV) & {\oldbf 16 $\pm$ 1} & {\oldbf 24 $\pm$ 3} & {\oldbf 8.2$^{+1.2}_{-0.9}$} & {\oldbf 10$^{+2}_{-1}$} & {\oldbf 13 $\pm$ 1} & {\oldbf 18 $\pm$ 2} \\
Norm$_{\rm n}$ ($\times$10$^{-4}$) & \nodata & \nodata & \multicolumn{2}{c}{\oldbf 0.8$^{+0.8}_{-0.4}$} & \nodata & \nodata \\
Tbabs $N_{\rm H}$ ($\times$10$^{22}$\,cm$^{-2}$)$^b$ & {\oldbf 9} & {\oldbf 11} & {\oldbf 9} & {\oldbf 10} & {\oldbf 10} & {\oldbf 11} \\
$\Gamma$ & \multicolumn{2}{c}{\oldbf 1.34$^{+0.05}_{-0.07}$} & \multicolumn{2}{c}{\oldbf 0.79$^{+0.09}_{-0.08}$} & \multicolumn{2}{c}{\oldbf 1.10 $\pm$ 0.06} \\
Normalization ($\times$10$^{-2}$) & {\oldbf 0.48$^{+0.04}_{-0.05}$} & {\oldbf 1.2 $\pm$ 0.1} & {\oldbf 0.38 $\pm$ 0.03} & {\oldbf 0.81$^{+0.08}_{-0.07}$} & {\oldbf 0.51 $\pm$ 0.04} & {\oldbf 1.09$^{+0.10}_{-0.09}$} \\
\tableline
Fe K$\alpha$ Energy (keV)$^b$ & \multicolumn{2}{c}{6.36} & \multicolumn{2}{c}{6.4} & \multicolumn{2}{c}{6.4} \\
Fe K$\alpha$ Width ($\sigma_{\rm Fe K\alpha}$)$^b$ & \multicolumn{2}{c}{0.1} & \multicolumn{2}{c}{0.1} & \multicolumn{2}{c}{0.1} \\
Normalization ($\times$10$^{-3}$\,photons cm$^{-2}$ s$^{-1}$) & {\oldbf $<$0.03} & {\oldbf 0.05 $\pm$ 0.03} & {\oldbf $<$0.03} & {\oldbf 0.047 $\pm$ 0.005} & {\oldbf $<$0.03} & {\oldbf 0.05 $\pm$ 0.02} \\
Fe K$\alpha$ EQW (eV) & {\oldbf $<$66} & {\oldbf 76$^{+11}_{-52}$} & {\oldbf $<$66} & {\oldbf 51$^{+30}_{-26}$} & {\oldbf $<$62} & {\oldbf 57$^{+24}_{-32}$} \\
Fe K$\alpha$ Flux ($\times$10$^{-13}$\,erg cm$^{-2}$ s$^{-1}$) & {\oldbf $<$1.9} & {\oldbf 3$^{+1}_{-2}$} & {\oldbf $<$1.7} & {\oldbf 3 $\pm$ 2} & {\oldbf $<$1.7} & {\oldbf 3$^{+1}_{-2}$} \\
\tableline
Absorbed Flux ($\times$10$^{-11}$\,erg cm$^{-2}$ s$^{-1}$)$^c$ & {\oldbf 2.33 $\pm$ 0.03} & {\oldbf 5.5 $\pm$0.1} & {\oldbf 2.33 $\pm$ 0.03} & {\oldbf 5.5 $\pm$ 0.1} & {\oldbf 2.29$^{+0.04}_{-0.05}$} & {\oldbf 5.5 $\pm$ 0.2} \\
Unabsorbed Flux ($\times$10$^{-11}$\,erg cm$^{-2}$ s$^{-1}$)$^d$ & {\oldbf 2.87 $\pm$ 0.04} & {\oldbf 7.0 $\pm$ 0.2} & {\oldbf 2.90 $\pm$ 0.04} & {\oldbf 7.0 $\pm$ 0.2} & {\oldbf 2.94 $\pm$ 0.06} & {\oldbf 7.1 $\pm$ 0.2} \\
\enddata
\tablecomments{\\*
$^a$ Cross-normalizations between detectors are calculated with respect to the value of FPMA at superorbital minimum. \\*
$^b$ The neutral hydrogen absorption column density, Fe K$\alpha$ line energy and width are frozen to the phase-averaged values at superorbital maximum. \\*
$^c$ Absorbed flux in the 3--10\,keV band. \\*
$^d$ Unabsorbed flux in the 3--10\,keV band.}
\label{Pulse Peak Spectral Parameters}
\end{deluxetable*}

\begin{figure}[h]
\centerline{\includegraphics[width=3.in]{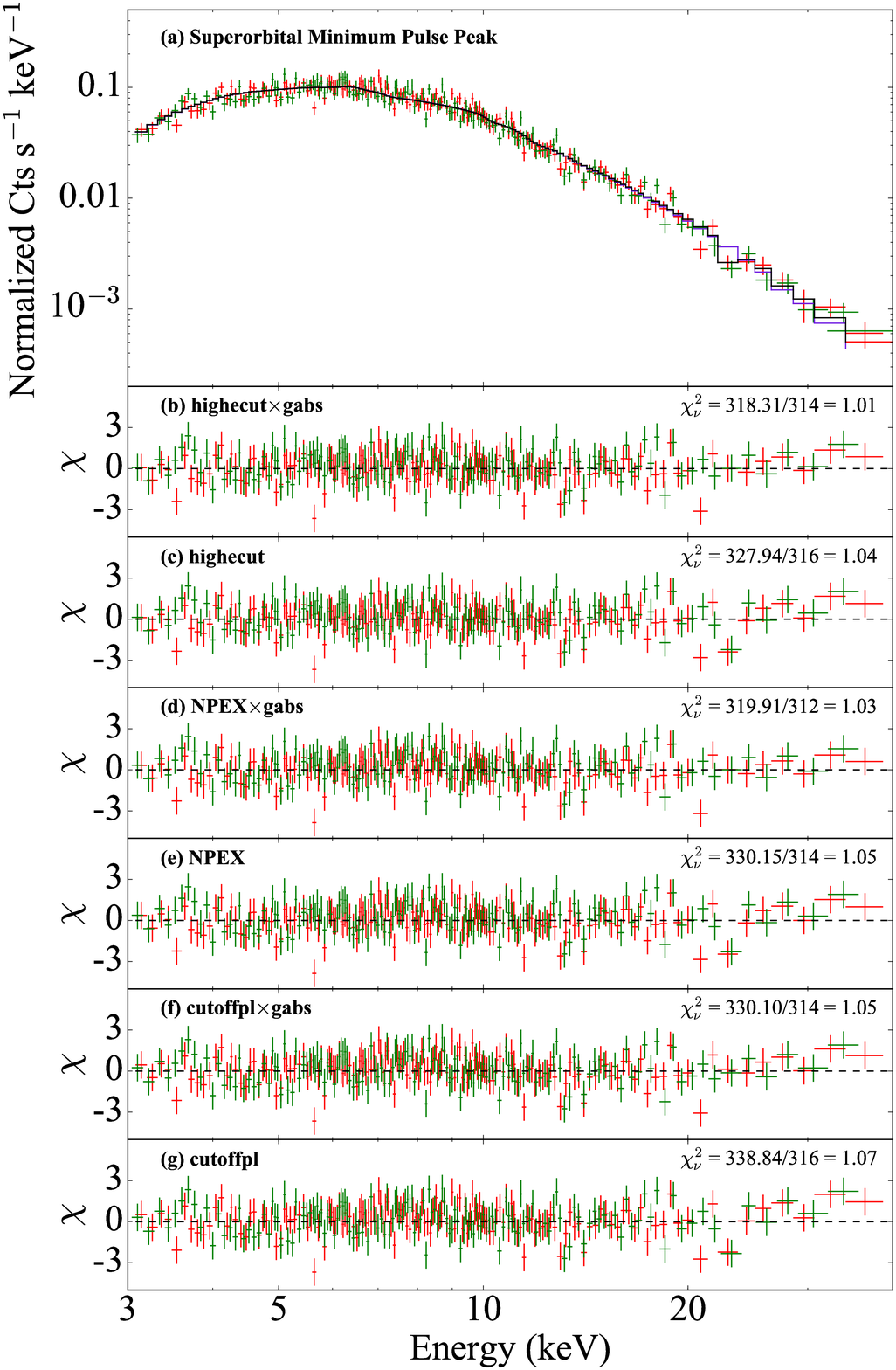}}
\figcaption[Jan8-2019_CompletePulsePeakMinimum.eps]{
(a) The pulse-peak \textsl{NuSTAR} spectra of IGR J16493-4348 at superorbital minimum.  The FPMA, and FPMB data are shown in red, and green, respectively.  The {\mybf \texttt{highecut}} model is shown in black.  (b) Residuals {\mybf for} the {\mybf \texttt{highecut}} model with the possible absorption line modeled by a Gaussian profile included in the model.  (c) Residuals for the {\mybf \texttt{highecut}} continuum model and Fe K$\alpha$ feature with no absorption line included in the model.  {\mybf (d) Residuals for the \texttt{npex} model with the possible absorption line modeled by a Gaussian profile included in the model.  (e) Residuals with no absorption line included in the \texttt{npex} model.  (f) Residuals for the \texttt{cutoffpl} model with the possible absorption line modeled by a Gaussian profile included in the model.  (g) Residuals with no absorption line included in the \texttt{cutoffpl} model.}
\label{Pulse Peak Spectrum}
}
\end{figure}

We detected possible negative residuals near $\sim$22\,keV in the pulse-peak spectrum at superorbital minimum (see Figure~\ref{Pulse Peak Spectrum}{\mybf (a)}).  {\mybf To investigate the possibility of an absorption feature, we only fit the pulse-peak spectrum at superorbital minimum and accounted for the residuals with a multiplicative Gaussian absorption feature.}  The addition of {\mybf an absorption line} reduces the $\chi^2/$d.o.f. {\mybf for the \texttt{highecut} model from 324.69/314 to 314.82/312 (see Figure~\ref{Pulse Peak Spectrum}(b)--(c))}.  We note the width of the feature cannot be constrained and instead we find the upper limit of the width to be 0.4\,keV.  The energy and optical depth of the possible absorption feature {\mybf for the \texttt{highecut} model} were found to be 23.1 $\pm$ 0.4\,keV and 0.8$^{+0.4}_{-0.2}$, respectively.  We do not find a significant change in the spectral parameters if instead we use a Lorentzian optical depth profile to describe the possible absorption line.  In this case, the $\chi^2/$d.o.f. {\mybf for the \texttt{highecut} model} is reduced from {\mybf 324.69/314 to 315.07/312}.

{\mybf We also investigated the possibility that the choice of the continuum influences the energy and shape of the possible absorption feature.  For the \texttt{npex} and \texttt{cutoffpl} models, the addition of an absorption line reduces the $\chi^2/$d.o.f. from 330.15/314 to 319.91/312 and 338.40/315 to 329.17/313, respectively (see Figure~\ref{Pulse Peak Spectrum}(d)--(g)).  The energy of the line is found to be 23.3\,$\pm$\,0.4\,keV and 23.3\,$\pm$\,0.5\,keV for the \texttt{npex} and \texttt{cutoffpl} models, respectively.}

To determine the significance of the inclusion of the possible absorption feature, we simulated 10$^{4}$ spectra using the Monte Carlo analysis described in \citet{2002ApJ...571..545P}.  Unlike our Monte Carlo simulations for Fe K$\alpha$ where we restricted the energy to that of the best fit model (see Sections~\ref{Phase-Averaged Spectral Analysis} and~\ref{Phase-Resolved Spectral Analysis}), we allowed the energy of the possible absorption feature to vary between 10\,keV and 40\,keV.  These are reasonable values for a possible CRSF \citep{2002ApJ...580..394C}.  We determined the probability of the possible absorption feature arising by chance to be 36.4$\%$ with 10$^{4}$ trials, {\oldbf which shows that the feature is not significant.}

\section{Discussion}
\label{Discussion}

\subsection{Spectral Evolution as a Function of Superorbital Period}
\label{Constraints on Accretion Geometry}

\begin{figure}[h]
\centerline{\includegraphics[width=3.in]{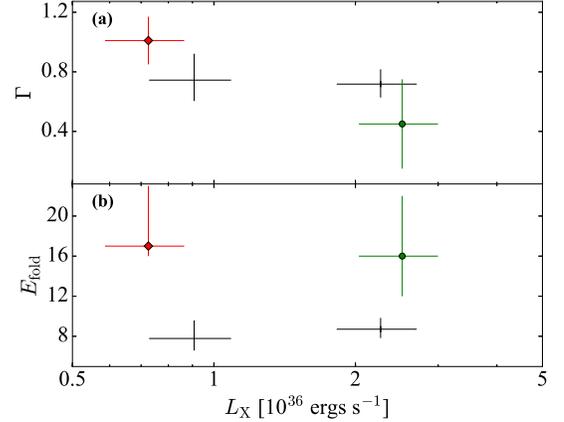}}
\figcaption[7-17-2018_DAiNPEXPearlman.eps]{
The evolution of the power law (a) photon index and (b) folding energy with the {\oldbf 1--10\,keV} X-ray {\oldbf luminosity} for the \texttt{npex} model.  The \textsl{Suzaku}/\textsl{Swift} BAT/\textsl{INTEGRAL} ISGRI \citep{DAi2011}, \textsl{Suzaku}/\textsl{Swift} BAT/\textsl{INTEGRAL} ISGRI \citep{DAi2011} and \textsl{NuSTAR} results (this work) are indicated by the green circles, red diamonds and black crosses, respectively.
\label{Coley D Ai Comparison}
}
\end{figure}

In Figure~\ref{Coley D Ai Comparison}, we plot the photon index and folding energy of the pulse-phase averaged \textsl{NuSTAR} and \textsl{Swift} XRT data as a function of the 1--10\,keV X-ray {\oldbf luminosity}.  For the \texttt{highecut} model, the pulse-phase averaged X-ray {\oldbf luminosity} of IGR J16493-4348 increased from {\oldbf (9\,$\pm$\,2)\,$\times$\,10$^{35}$\,erg s$^{-1}$ at superorbital minimum to (2.4\,$\pm$\,0.4)\,$\times$\,10$^{36}$ erg s$^{-1}$ at superorbital maximum (see Section~\ref{Phase-Averaged Spectral Analysis}).}  To place the data into context, we also show the change in the photon index and folding energy with respect to 1--10\,keV X-ray {\oldbf luminosity} of the broadband \textsl{Swift} BAT and the \textsl{INTEGRAL} ISGRI, together with pointed \textsl{Swift} XRT and \textsl{Suzaku} observations reported in \citet{DAi2011}.  The \textsl{Swift} XRT and \textsl{Suzaku} observations of IGR J16493-4348 took place at MJD\,53,805.9--53,806.4 and MJD\,54,013.9--54,014.4, respectively; which correspond {\oldbf to} superorbital phases $\sim$0.05--0.07 and $\sim$0.42--0.45 or slightly earlier superorbital phases than our \textsl{NuSTAR} campaign.  Their X-ray {\oldbf luminosities,} however, are similar to our \textsl{NuSTAR} observations (see Figure~\ref{Coley D Ai Comparison}).

In accreting X-ray pulsars, the shape of the pulse profiles {\oldbf has been} found to depend on the emission processes and the relative contribution of {\oldbf the} two accretion columns {\mybf \citep[][Falkner et al. 2019]{1985ApJ...299..138M,1989ESASP.296..433K}}.  Our \textsl{NuSTAR} observations of IGR J16493-4348 show that despite the increase in X-ray flux, the pulse profiles show no significant changes in shape between superorbital minimum and maximum.  The pulse profiles in both observations were found to weakly depend on energy and be double-peaked in structure (see Figure~\ref{Energy Resolved Spin Profiles}).  {\oldbf This may indicate} that the emission properties in the accretion column may not change between the two superorbital phases \citep[e.g. A 0535+26{\oldbf ,}][]{2017A&A...608A.105B}.  Using \textsl{RXTE} PCA data that span {\oldbf times} between MJD\,55,843.1 and MJD\,55,852.6, corresponding to superorbital phases 0.62--0.09, \citet{2019ApJ...873...86P} also found a double-peaked shape of the pulse profiles and a weak energy dependence.  They also observed the pulsed fraction to increase with increasing energy, which we confirm with \textsl{NuSTAR} (see Figure~\ref{NuSTAR Pulse Fraction}).  We suggest the similar properties of pulse profiles as seen by both \textsl{NuSTAR} and \textsl{RXTE} may {\oldbf be} linked to an accretion regime that does not change between superorbital minimum and maximum {\oldbf \citep[see][and references therein]{2015MNRAS.452.1601P}.}

In X-ray binaries that host accretion powered pulsars, the physical conditions inside the accretion column depend on the mass accretion rate.  The resulting X-ray emission can be characterized in terms of the local Eddington limit \citep[$L_{\rm crit}$,][]{2012A&A...544A.123B}, which for a magnetic dipole geometry is proportional to the magnetic field strength.  {\oldbf As indicated in Section~\ref{Introduction}, the magnetic field of the neutron star is often directly measured by cyclotron resonant scattering features (CRSFs).  We do not find any significant CRSFs in the pulse phase-averaged or phase-resolved \textsl{NuSTAR} spectra of IGR J16493-4348, which is a possible indication that the magnetic field might be predominately seen under small viewing angles \citep{2017A&A...601A..99S}}.

{\oldbf At low to intermediate accretion rates, the X-ray luminosity is below the critical value where radiation presssure becomes important.  The in-falling matter may be primarily decelerated by Coulomb interactions, and is thought to form an accretion mound close to the surface of the neutron star \citep[][and references therein]{2015MNRAS.452.1601P}.  Our \textsl{NuSTAR} results show that the X-ray luminosity observed in IGR J16493-4348 is on the order of 10$^{35}$--10$^{36}$\,erg s$^{-1}$, which is consistent with this picture \citep[see][and references therein]{2012A&A...544A.123B}.}

{\oldbf The spectral shape in X-ray binary pulsars that are accreting at low to intermediate accretion rates is observed to harden towards higher X-ray fluxes \citep[e.g. Her X-1; A 0535+26,][]{2011A&A...532A.126K,2017A&A...608A.105B}.  Our pulse-phase resolved \textsl{NuSTAR} results of IGR J16493-4348 near the broad main peak of the pulse profile show that the pulse-phase resolved hardness ratio increases between superorbital minimum and maximum (see Figure~\ref{Pulse Resolved Max Spectral Fits}(b)), which is a possible indication that the spectrum hardens with increasing X-ray flux.  This is also supported by the ratio of the two peak-spectra, which shows a clear slope indicating the increasing dominance of the flux at superorbital maximum toward higher energies (see Figure~\ref{Pulse Peak Ratio}(b)).}

{\mybf It is interesting to note that \citet{DAi2011} found the photon index flattens between superorbital minimum and maximum (see Figure~\ref{Coley D Ai Comparison}(a)).  While this may also suggest that the spectral shape of IGR J16493-4348 hardens with increasing X-ray flux, the photon indices observed with \textsl{NuSTAR} were found to be consistent between the superorbital minimum and maximum.  As shown in Figure~\ref{Coley D Ai Comparison}, they were additionally found to be consistent with those reported in \citet{DAi2011}.  It should also be noted that \citet{DAi2011} found the folding energy in their broadband BAT and ISGRI, together with pointed \textsl{Swift} XRT and \textsl{Suzaku} observations to be systematically higher than those we derived with \textsl{NuSTAR} (see Figure~\ref{Coley D Ai Comparison}(b)).  These differences may result from the fact that the spectral shape may have changed between the observations reported in \citet{DAi2011} and our \textsl{NuSTAR} observations.  Another possible reason is the energy gap between the soft (\textsl{Swift} XRT and \textsl{Suzaku}) and hard (\textsl{INTEGRAL} and \textsl{Swift} BAT) bands, which may affect the spectral fits reported in \citet{DAi2011}.}

This {\oldbf observed} spectral hardening seen near the main peak of the pulse profile could possibly be explained by {\oldbf Compton-saturated emission from the sidewall of the optically thick accretion column \citep{2015MNRAS.452.1601P}.}  We found the folding energy near the main pulse peak to increase between superorbital minimum and maximum, but no change in the photon index was found (see Table~\ref{Pulse Peak Spectral Parameters}).  This may suggest the average temperature of the Comptonizing gas increases between superorbital minimum and maximum.  Due to this temperature increase, the plasma more efficiently upscatters photons to higher energies via the inverse Compton effect, resulting in a harder observed spectrum as is observed in the pulse-peak superorbital maximum spectrum.

\subsection{Comparison with 2S 0114+650}

To place IGR J16493-4348 in context with other wind-fed SGXBs where superorbital variability is found, we compare it with 2S 0114+650.  2S 0114+650 is a wind-fed SGXB where a $\sim$30.7\,day superorbital period was found \citep{2006MNRAS.367.1457F}.  The spectral type of the mass donor in 2S 0114+650 was found to be B1 I \citep{1996A&A...311..879R}, which is similar to the B0.5 Ia spectral type in IGR J16493-4348 \citep{2019ApJ...873...86P}, and its distance was estimated to be ∼7.2 kpc.

We first discuss the \textsl{Swift} BAT observations of IGR J16493-4348 in comparison to the long-term monitoring of the 30.7\,day cycle present in 2S 0114+650. Using \textsl{RXTE} ASM data that spanned $\sim$8.5\,yr, \citet{2006MNRAS.367.1457F} found that the amplitude of the 30.7\,day modulation changed as a function of time.  This was recently confirmed by \citet{2017ApJ...844...16H} using ASM and BAT data spanning $\sim$20\,years.  No significant changes in its frequency were found, which is {\mybf similar to what we find for IGR J16493-4348.}

We also discuss pointed observations of 2S 0114+650 and how they compare with our joint \textsl{NuSTAR} and \textsl{Swift} campaign for IGR J16493-4348. In their \textsl{RXTE} campaign, which covered two cycles of the $\sim$30.7\,day period, \citet{2008MNRAS.389..608F} found no changes in the intrinsic neutral column density on superorbital timescales.  The spectral shape in 2S 0114+650 was found to harden as the superorbital cycle progressed from minimum to maximum \citep{2008MNRAS.389..608F}, which is similar to our pulse-peak analysis of IGR J16493-4348 (see Sections{\oldbf ~\ref{Phase-Resolved Spectral Analysis}--\ref{Pulse-peak Spectral Analysis}}).  We note \citet{2008MNRAS.389..608F} have shown that the photon index increases by a factor of two between superorbital maximum and minimum in 2S 0114+650, which is not seen in {\mybf our observations of} IGR J16493-4348.

Although both 2S 0114+650 and IGR J16493-4348 are both mediated by wind accretion, their superorbital modulations may show somewhat different spectral behavior.  The spectrum of 2S 0114+650 significantly hardened towards higher luminosities, but the correlation between spectral hardness and X-ray luminosity {\oldbf in IGR J16493-4348} may be weaker {\oldbf and is observed only near the peak of the pulse profile (see Sections{\oldbf ~\ref{Phase-Resolved Spectral Analysis}--\ref{Pulse-peak Spectral Analysis}})}.  To investigate these possible differences between 2S 0114+650 and IGR J16493-4348, we compare the accretion regimes between the two sources.

\citet{2008MNRAS.389..608F} found that the average absorbed 3--50\,keV X-ray flux of 2S 0114+650 was 2.3$\times$10$^{-10}$\,erg\,cm$^{-2}$\,s$^{-1}$ and the fully covered absorption to be 3.2$^{+0.9}_{-0.8}$$\times$10$^{22}$\,cm$^{-2}$.  To investigate if IGR J16493-4348 and 2S 0114+650 are accreting in similar accretion regimes, we calculated the unabsorbed X-ray luminosity in 2S 0114+650.  Assuming the spectral parameters reported in \citet{2008MNRAS.389..608F}, we corrected for absorption using \texttt{PIMMS} and found the unabsorbed 3--50\,keV X-ray flux in 2S 0114+650 to be 2.4$\times$10$^{-10}$\,erg\,cm$^{-2}$\,s$^{-1}$.  The average 3--50\,keV X-ray luminosity of 2S 0114+650 is found to be 1.5$\times$10$^{36}$\,erg\,s$^{-1}${\oldbf , which is {\mybf of the same order of magnitude} to the average 3--50\,keV X-ray luminosity observed in IGR J16493-4348 with \textsl{NuSTAR} and \textsl{Swift} XRT (see Section~\ref{Phase-Averaged Spectral Analysis}).  {\oldbf Our results near the peak of the pulse profile show that changes in the spectral shape of IGR J16493-4348 are similar to those of 2S 0114+650, albeit the trend in IGR J16493-4348 is somewhat weaker.}

{\mybf \subsection{Superorbital Modulation in Ultraluminous X-ray Sources}}

{\mybf Superorbital modulation on timescales of tens of days has also been detected in Ultraluminous X-ray (ULX) Pulsars \citep[e.g. NGC 5907 ULX1, NGC 7793 P13, M82 X-2;][]{2016ApJ...827L..13W,2018A&A...616A.186F,2019ApJ...873..115B}, accreting neutron stars with apparent luminosities in excess of 10$^{39}$\,erg s$^{-1}$.  While such timescales are similar to those seen in X-ray binaries accreting at sub-Eddington rates, it is important to note that the timing and spectral properties of ULXs show significant differences compared to those observed in wind-fed SGXBs such as IGR J16493-4348.  Due to their super-Eddington X-ray luminosities, the mode of accretion in ULX pulsars has been ascribed to Roche-lobe overflow \citep{2014Natur.514..202B} and the superorbital mechanism is likely to be partially driven by a precessing accretion disk \citep[see][and references therein]{2017ApJ...834...77F}.  The modulation amplitude between superorbital minimum and maximum in ULX pulsars show similarities with XRBs where superorbital variations are driven by a precessing disk.  For example, the amplitude of the 60\,day modulation observed in M82 X-2 was found to vary by two orders of magnitude \citep{2019ApJ...873..115B}, which is similar the variability observed in LMC X-4 \citep{2015AstL...41..562M}.

Some ULX pulsars have also been observed to exhibit ``off" states, where their X-ray fluxes were found to be up to several orders of magnitude lower than expected from an extrapolation of their observed periodic signals \citep{2016ApJ...827L..13W}.  While the amplitude of the superorbital modulation was found to significantly change on long timescales (see Figure~\ref{Dynamic Power Spectrum}), no evidence of ``off" states was revealed in our \textsl{Swift} BAT observations in IGR J16493-4348.}

\subsection{Superorbital Modulation Mechanism}
\label{Mechanisms}

\subsubsection{Precessing Accretion Disk}
\label{Precessing Accretion Disk}

We first discuss the possibility that the $\sim$20\,day modulation in IGR J16493-4348 is driven by a precessing accretion disk \citep[e.g. Her X-1;][]{2000ApJ...539..392S}.  Large variations in the intrinsic neutral hydrogen absorption column are found {\mybf in this case,} \citep[e.g. Her X-1;][]{2002MNRAS.337.1185R} resulting in sharp dips in the superorbital profile \citep[e.g. Her X-1; SMC X-1,][]{2006AstL...32..804K,2007ApJ...670..624T}.  {\mybf No significant changes in $N_{\rm H}$ are observed between superorbital minimum and maximum (see Table~\ref{Broadband Spectral Parameters}), providing evidence against this model.  Dramatic changes in the strength of X-ray pulsations have also been found in sources where superorbital modulation is linked to the precession of an accretion disk such as SMC X-1 \citep{2019ApJ...875..144P} and ULX pulsars \citep[e.g. M82 X-2,][]{2014Natur.514..202B}.  In these sources, the variations in the strength of the pulsations were not accompanied by large changes in X-ray flux.  No such changes in the pulse profile and pulsed fraction that can be explained by variations in absorption are observed in IGR J16493-4348, which suggests that the 20\,day modulation is probably not driven by a precessing accretion disk.}

If the $\sim$20.6\,day modulation was driven by a moderately long-lived prograde transient accretion disk, long-term variations in its modulation amplitude may be observable \citep[e.g. 2S 0114+650;][]{2017ApJ...844...16H}.  In the BAT dynamic power spectrum, we found a low amplitude in the superorbital modulation spanning $\sim$600\,days (see Figure~\ref{Dynamic Power Spectrum}), which is similar to the formation and dissipation timescale of a transient accretion disk {\oldbf proposed} to be present in 2S 0114+650.  We note in systems where a transient disk may be present, the neutron star is expected to rapidly spin up due to the large angular momentum transfered to it \citep[e.g. OAO 1657-415;][]{2012ApJ...759..124J}.  \citet{2019ApJ...873...86P} found no evidence of a rapid spin up torque in their pulsar timing analysis using the \textsl{RXTE} PCA, which {\mybf suggests that a transient accretion disk may not be present.}

\subsubsection{Stellar Triple System}

Next, we discuss the possible case that IGR J16493-4348 is part of a triple-star system \citep[e.g. 4U 1820-30;][]{2001ApJ...563..934C}.  In a triple-star system, the eccentricity of the inner binary is modulated at a long term period by tidal forces of a third companion star orbiting the center of mass between it and the inner binary \citep{1979A&A....77..145M}.  This period is inversely proportional to the period of the inner binary, and directly proportional to the orbital period of the third companion \citep[see Equation 19 in][]{2007MNRAS.377.1006Z}.  If this model is applied to IGR J16493-4348, we calculate the third period to be 11.666$\pm$0.002\,days.  {\oldbf This period is only a factor of $\sim$1.7 times larger than the binary orbital period measured with \textsl{RXTE} and \textsl{Swift} \citep{2019ApJ...873...86P}, which may imply an unstable orbital configuration if IGR J16493-4348 were part of a triple star system.}

The stability of a triple star system depends on the ratio between the semimajor axis of the third companion star and the orbital separation of the components in the inner binary \citep{2007MNRAS.377.1006Z,2008msah.conf...11M}.  Combining the ratio between the outer and inner periods with Kepler's third law, we calculate the ratio between the outer and inner semimajor axes to be $\sim$1.4.  This close configuration may result in perturbations of the binary motion on time-scales between the orbital period and the superorbital modulation, which were not observed with \textsl{RXTE} or \textsl{Swift} \citep[see e.g.][]{2019ApJ...873...86P}. {\mybf It is worth noting, depending on the mass of the third companion star, that strong perturbations from a binary orbit may be detectable in pulsar timing residuals \citep[e.g. PSR J0337+1715,][]{2014Natur.505..520R}.  These perturbations were not found in the \textsl{RXTE} pulsar timing residuals reported in \citet{2019ApJ...873...86P}, providing additional evidence against a third companion star.}

\subsubsection{Precession of the Donor Star}

{\oldbf If} a precessing donor star surrounded by an equatorially enhanced wind were the cause of the $\sim$20.06\,day period in IGR J16493-4348, long-term changes in the neutral hydrogen absorption column density may be observed \citep[e.g. GX 304-1,][]{2017MNRAS.471.1553K}.  We do not find any significant variations in $N_{\rm H}$ between superorbital minimum and maximum.  {\mybf However, we} caution against ruling {\oldbf out} a precessing equatorial wind since the absorbing material might not be along the line of sight.

\subsubsection{Corotating Interaction Regions in the Stellar Wind}

An alternative possibility is that the $\sim$20.6\,day cycle could be driven by large-scale corotating interaction regions (CIRs) in the wind of the B0.5 Ia donor star \citep{2017A&A...606L..10B}.  Changes in the mass accretion rate may be partially modulated by the interaction between the neutron star and the CIRs.  Phase-locked flares that were possibly attributed to large-scale structures in the wind of the donor star have been identified in the SFXT IGR J16479-4514 \citep{2013MNRAS.429.2763S}, which also shows strong superorbital modulation.

In a non-synchronous rotating binary, the angular velocities of the neutron star and the CIR would be different \citep[see][and references therein]{2017A&A...606L..10B}, resulting in a beat period on superorbital timescales.  \citet{2017A&A...606L..10B} applied this model to IGR J16493-4348 and found that a single CIR with a period of $\sim$10.3\,days could explain the $\sim$20.06\,day superorbital period.  We note the $\sim$20.6\,day modulation is persistently detected in the dynamic power spectrum spanning a timescale of more than 12\,years (see Figure~\ref{Dynamic Power Spectrum}(a)).  {\mybf This suggests that, if CIRs are the cause of the superorbital modulation in these systems, they would have to be stable over timescales of several years \citep{2019ApJ...873...86P}}.

\subsubsection{Tidal Oscillations}

Finally, we discuss the possible case that the superorbital modulation in IGR J16493-4348 is driven by a non-synchronously rotating donor star \citep{2005A&A...437..641M,2006A&A...458..513K}, which could exhibit several different periodicities due to tidal oscillations \citep{1977A&A....57..383Z,2005A&A...437..641M}.  Such oscillations could produce a localized structured wind, which would drive periodic modulation in the X-ray band when accreted onto the neutron star.  The period of these oscillations was calculated to be on superorbital timescales for a circular orbit.  From a pulsar timing analysis, \citet{2019ApJ...873...86P} {\mybf showed that the binary is likely in a nearly circular orbit}, which meets the requirement of the tidal oscillation model.

\section{Summary and Conclusions}
\label{Conclusion}

In this paper, we have presented two \textsl{NuSTAR} observations of IGR J16493-4348, which coincide with the minimum and maximum of one cycle of its $\sim$20\,day superorbital modulation, and long-term observations of the superorbital period by \textsl{Swift} BAT.  An analysis of the BAT data using the dynamic power spectra and fractional root mean square methods reveals strong variations in the amplitude of the superorbital modulation, but we do not observe changes in the period.  The fractional rms of the $\sim$20.06\,day period closely tracks the peak power, providing additional evidence that its amplitude significantly changes with time.

Our results indicate the neutron star rotation period is consistent with that reported by \citet{2019ApJ...873...86P} at the 1$\sigma$ confidence interval.  This {\oldbf suggests that no significant long-term neutron star rotation period derivative was detected between the \textsl{RXTE} and \textsl{NuSTAR} observations.}  No significant changes in the 3--50\,keV pulse profiles between the two observations are found, which suggests a similar accretion regime at superorbital minimum and maximum.

We have presented a pulse phase-resolved spectral analysis of IGR J16493-4348 for the first time.  Our results show that while the joint \textsl{NuSTAR} and \textsl{Swift} XRT pulse phase-averaged spectral continuum revealed no significant changes between superorbital minimum and maximum, we observe possible evidence of luminosity-dependent variability in the pulse phase-resolved spectra.  We found the spectral shape near the broad main peak of the pulse profile hardens between superorbital minimum and superorbital maximum, which is consistent with the subcritical accretion regime.  It may {\mybf be} possible that the spectral hardness evolution seen in IGR J16493-4348 could be explained by thermal Comptonization in a collisionless shock model \citep[e.g. Cep X-4;][]{2017A&A...601A.126V}.

We also found a weak emission line at 6.4\,keV at superorbital maximum, but it is not significant at superorbital minimum.  The origin of the 6.4\,keV emission line is due to neutral Fe or Fe in a low ionization state, which is present in many {\mybf X-ray binaries}.  Our pulse-phase resolved analysis indicates that the flux of the Fe K$\alpha$ line does not track the pulse profile, a possible indication that the region responsible for the Fe K$\alpha$ emission is not close to the neutron star.

Our \textsl{NuSTAR} and \textsl{Swift} analysis shows that while the mechanism responsible for the superorbital modulation remains elusive, we can now {\oldbf begin to} constrain it.  Mechanisms where we might expect a significant change in the neutron hydrogen column density -- a precessing accretion disk, a precessing equatorially enhanced wind -- are unlikely.  A transient accretion disk {\mybf may also be unlikely} since the neutron star shows no {\mybf indications of a} rapid spin-up torques \citep{2019ApJ...873...86P}.  A triple-star system {\oldbf is unlikely} since the period of the third object is calculated to be 11.666$\pm$0.002\,days, which {\oldbf may lead to an unstable orbital configuration.}

The superorbital dependence of the spectral shape in IGR J16493-4348 particularly near the broad main peak of the pulse profile shows similarities to 2S 0114+650, which is the prototypical wind-fed SGXB exhibiting superorbital modulation \citep{2006MNRAS.367.1457F}.  While the spectral shape in both sources each hardens from superorbital minimum to maximum, an anticorrelation between photon index and X-ray luminosity is only observed in 2S 0114+650.  The behavior of spectral hardness in both sources; however, may suggest that the superorbital mechanism is linked to a variable accretion rate.  Superorbital mechanisms that explain the variable accretion rate such as tidal oscillations or large structures in the donor star wind remain possible.

To further understand the mechanism responsible for the $\sim$20.06\,day superorbital cycle in IGR J16493-4348, additional multi-wavelength observations are required.  The study would benefit from optical/near-infrared photometry, which may confirm or preclude variations in the donor star or its wind as the driving mechanism of the superorbital modulation.

\acknowledgements

{\mybf We thank the anonymous referee for useful comments.}  We {\mybf also} thank Drs. Patricia Boyd, Sebastian Falkner, Illeyk El Mellah and Enrico Bozzo for useful discussion {\oldbf and the \textsl{NuSTAR} Operations, Software and Calibration teams for scheduling and the execution of these observations.}  A. B. Pearlman acknowledges support by the Department of Defense (DoD) {\mybf through the National Defense Science and Engineering Graduate (NDSEG) Fellowship Program and by the National Science Foundation (NSF) Graduate Research Fellowship under Grant No. DGE-1144469}.  This research has made use of the XRT Data Analysis Software (XRTDAS) developed under the responsibility of the ASI Science Data Center (ASDC), Italy and the \textsl{NuSTAR} Data Analysis Software (NuSTARDAS) jointly developed by the ASI Science Science Data Center (ASDC, Italy) and the California Institute of Technology.  We thank NASA's {\mybf 14-ADAP14-0167 grant and} \textsl{NuSTAR} Guest Observer Grant 14-NUSTAR14-0007 for support.



\end{document}